\documentclass[aps,showpacs,eps,preprint]{revtex4}
\usepackage{amssymb}
\usepackage{amsmath}
\usepackage{epsf}
\usepackage{mathrsfs}
\usepackage[dvips]{graphics}
\usepackage[dvips,usenames]{color}
\begin{document}
\def \beq{\begin{equation}}
\def \eeq{\end{equation}}
\def \bea{\begin{eqnarray}}
\def \eea{\end{eqnarray}}
\def \bem{\begin{displaymath}}
\def \eem{\end{displaymath}}
\def \f{\frac{\Omega}{\omega_\perp}}
\def \fw{\Phi_{\omega}}
\def \fl{\Phi_{\Omega}}
\def \P{\Psi}
\def \Po{\overline{\Psi}}
\def \p{\psi}
\def \hp{\hat{\psi}}
\def \hpa{\hat{\psi}^{\dagger}}
\def \bs{\boldsymbol}
\title{Vortices in Atomic Bose-Einstein Condensates: an Introduction}
\author{Sankalpa Ghosh}
\affiliation{ Physics Department, Technion I.I.T. Haifa-32000, Israel}
\date{\today}

\begin{abstract}
The occurrence of vortices in atomic Bose-Einstein condensates (BEC) 
enables a description of their superfluid behaviour. 
In this article we present a pedagogical introduction to the 
vortex physics in  trapped atomic BECs.  
The mechanism of the vortex nucleation
in an atomic BEC is discussed in detail.
We also discuss a recently proposed approach 
which treats the problem of vortex nucleation 
using a one-particle  Schr$\ddot{o}$dinger equation with 
non-local and chiral boundary conditions.
\end{abstract}
    
\pacs{03.75.Kk, 03.75.Lm, 05.30.Jp, 67.40.Db}
\maketitle
\tableofcontents

\section{Introduction}
One of the prominent features of a superfluid is the way
it behaves under rotation \cite{Leggett}.   
The second derivative of the free energy with respect to the rotational
frequency is proportional to the superfluid density and  vanishes
for a normal fluid. 
Therefore a way to distinguish between superfluids and normal fluids is
through  their response to rotation. 
In contrast to a normal fluid, which rotates like a
rigid body at thermal equilibrium, the thermodynamically stable state
of a superfluid does not rotate at low enough frequency. 
At higher frequencies the angular momentum appears as 
vortex filaments at which
the superfluid density vanishes. The circulation of the velocity field 
flow 
around a closed contour which encircles the vortex, is
quantized \cite{On49, Feyn55}.
This is a consequence of the
properties of the macroscopic wavefunction, whose phase changes
by an integer multiple of $2\pi$ around the vortex filaments.
The atomic Bose-Einstein
condensates (BEC) \cite{book, book1, Dalf99, CastinR}
provide a system for which this superfluid behaviour 
can be studied in the weak-coupling regime. 

In this article we present a  detailed introduction to 
the problem of vortex nucleation in atomic Bose-Einstein
condensates which includes the recent developments in  
this field following the experimental detection of vortices 
in BEC 
\cite{JILA1, JILA2, MCWD99, MCWD00, CMD00, CMBD01, MCBD01, MIT1, MIT2, OXFORD}.
Part of the contents overlaps with those of a recent review  
by Fetter and Svidzinsky\cite{fsr01}
which is a much more broad based account of vortices in BEC in general.
However we emphasized here the problem of vortex nucleation and provide
also a detailed background theoretical framework aiming a more general
readership.

The article is organized as follows
In the first few sections we described in detail the vortices
in trapped atomic condensates, the associated quantization of circulation, 
the thermodynamic stability of a vortex 
and the evaluation of
the characteristic nucleation frequency of a vortex based on thermodynamic
arguments.
In Section II, we focus on  a single boson problem in a harmonic
confinement which has been detailed in a number of  
references (for example see chapter 6, ref. \cite{BR85}). We discuss 
the results in a static as well as in a rotating reference frame. 
The rotating trap problem is important in view of the fact that 
the experiments on vortices in 
atomic condensates are generally performed in a rotating trap.
We show why a vortex cannot be 
nucleated in a mechanically stable rotating harmonic trap if the bosons
are non-interacting 
whereas a finite amount of interaction makes vortex states energetically 
feasible. In Section III, we describe the Bogoliubov theory
which deals with weakly interacting bosons at very low temperature
\cite{Bogo47} (see \cite{Fetter72} for a review and further 
developments). 
The extension of this theory to the trapped 
atomic condensate has been studied in a number of references 
\cite{Mark96, Fett96, Dodd97, Rokh97, FR98, SF98, Tomo99, Mach99, Tomo01, 
ALF98, SMS} which includes
the cases of static as well as the rotating traps and has been
reviewed in \cite{Dalf99, CastinR, fsr01}. We  show how
the Bogoliubov theory leads to the  
Gross-Pitaevskii (GP) equation \cite{Gros61, Pita61} which is very successful
in explaining the properties of the condensate 
at very low temperature as observed in recent 
experiments, both with and without vortices. 
We also discuss the meaning of the condensate wavefunction
\cite{Feyn54, ady71} and how it is related to superfluidity
(see chapter {\rm 5} of \cite{Noz}).
We then explain how the vortex solutions
of the GP equation lead to the quantization of circulation, 
an idea which dates back to L. Onsager and 
R. P. Feynman \cite{On49, Feyn55, Don91}.

This is followed by 
a discussion showing how  
the characteristic frequency of vortex nucleation in a trapped 
condensate can be found from the solution of the GP equation 
\cite{Ginz58, Pethick96, sd96, dps96, BR99}. The main aspect of the 
Bogoliubov theory is the description of the weakly interacting 
bosons in terms of 
a condensate wavefunction and of non-interacting quasiparticles
known as the Bogoliubov excitations. We discuss these 
Bogoliubov excitations and show how they determine the stability of 
a solution of the Gross-Pitaevskii equations. We emphasize 
the instability of a vortex solution in a static trap 
\cite{Dodd97, Rokh97, ALF98, SF00L} 
and clarify how this instability is removed when the trap is put into
rotation \cite{Tomo01, SF00L, LF99}. We stress that this
aspect explains 
why vortices are not thermodynamically stable below a characteristic 
rotational frequency of the trap \cite{fsr01, LF99}.

In Section V, we describe the Thomas-Fermi (TF) approximation 
\cite{Pethick96, sd96} which gives analytical expressions for experimentally
measurable quantities while has been very successful in 
explaining the behaviour of a large condensate. We also discuss the correction 
to the TF approximation required to describe the condensate profile
at the boundary.  
We describe  
the condensate profile with and without  vortex under this approximation
and mention how an analytic expression for the vortex nucleation can be 
derived under this approximation \cite{Lund97, Sinh97, F99}. We then 
turn to the study of the collective excitations of the condensate within the
TF approximation \cite{Sinh97, SF98}. For this we introduce
the hydrodynamic description of the condensate  
and show how it leads to the dispersion law of the collective excitations 
\cite{SS96}. We  discuss in more detail the quadrupole
mode \cite{SZ98}
since it plays an important role in the
process of vortex nucleation. 

In section VIII, we discuss experiments  
experiments \cite{JILA1, JILA2, MCWD99, MCWD00, CMD00, 
CMBD01, MCBD01, MIT1, MIT2, OXFORD}
and compare them 
with the theoretical predictions.
We mention how the 
experimental observations  point out the limitation of 
the thermodynamic critical frequency of vortex nucleation and points 
towards a more local theory of vortex nucleation which we discuss in the 
next few sections. Vortices are nucleated from the surface. So
to construct such a local theory one needs to know 
the condensate profile and its collective excitations at the surface
of the trap. To that purpose we first account for the correction 
to the TF approximation required to describe the condensate profile
at the boundary \cite{dps96, DLS96, FF98}. Then we discuss how this correction 
is included into the hydrodynamic theory to find out the modified
dispersion relation of the surface modes \cite{AKPS99, DS00, JA01}. 
Once the dispersion relation of the surface modes are identified 
using the Landau criterion \cite{LandauC}, it is possible to determine 
the rotational frequency for which a given surface excitation becomes unstable
towards the vortex nucleation \cite{DS00, JA01, SMS}. 
We particularly emphasize
that the characteristic frequency determined in this way 
\cite{JA01} agrees very well with the value 
experimentally observed \cite{MIT1}.

In section X, we discuss a different approach we have proposed  
\cite{AG03} to study the 
problem of vortex nucleation where one replaces the non-linear interaction
term in the GP equation
by a set of non-local and chiral boundary conditions. The problem 
we study is strictly two-dimensional. Such boundary conditions give 
a natural splitting of the Hilbert space of the problem into bulk and edge 
states (which are two-dimensional analogues of the surface states). We show
that the characteristic frequency of the vortex nucleation determined 
in this way is very close to the values experimentally observed \cite{MIT1}. 
In experiments a rotated trap is deformed in the plane of the rotation
\cite{CMBD01}.
If the deformation in the trap is not very large, it can be  
associated with 
a quadrupolar density deformation and the corresponding velocity
field. The dynamics of this quadrupolar mode in the rotating frame 
determines the vortex  nucleation in the condensate
by overcoming a surface barrier. This issue has been investigated in detail
both experimentally \cite{CMBD01, MCBD01, OXFORD}  
as well as theoretically \cite{RZS01, CS01, KPSZ01} which we also 
discuss briefly. 
We conclude 
our article by summarizing these discussions. 

\setcounter{equation}{0}%
\renewcommand{\theequation}{\mbox{\arabic{section}.\arabic{equation}}}%
\section{Single-boson problem}\label{subsb}
We start our discussion by listing the results from 
the solutions of the  time-independent 
Scrh$\ddot{o}$dinger equation of a trapped boson \cite{BR85}), namely, 
\beq H_0 \Phi(\bs{r})~=~\left(-\frac{\hbar^2}{2m} 
+ V_{tr}(\bs{r}) \right)\Phi(x,y,z)~=~E\Phi(x,y,z) \label{sce} \eeq
The trap can be either static or rotating. 
In the later case 
the hamiltonian in the lab-frame is  
time-dependent. A time-independent hamiltonian  ($H_{rot}$)
is obtained by going 
into the co-rotating frame of the trap which is related to the static 
hamiltonian in the lab frame ($H_{0}$) by  
\beq H_{rot}=H_{0} - \bs{\Omega}\cdot \bs{L} \eeq

\subsection {Anisotropic case}\label{ANSO} 
If the trap is harmonic 
the Schr$\ddot{o}$dinger equation (\ref{sce}) rewrites  
\beq \left[-\frac{\hbar^2}{2m}\nabla^2 + \frac{m}{2}(\omega_{x}^2 x^2 
+ \omega_y^2 y^2 + \omega_z^2 z^2) \right ]\Phi(x,y,z)=E \Phi(x,y,z) 
\label{se}\eeq
The most general case is one in which all the trap frequencies 
are different. 
The problem is separable in Cartesian coordinates, namely, 
$\Phi(x,y,z)=\phi_{1}(x)\phi_2 (y) \phi_3(z)$. 
The eigenfunctions 
and the energy eigenvalues are respectively those 
of a three dimensional harmonic oscillator and are defined by  
a set of three quantum numbers $\bs{n} =\{n_x, n_y, n_z\}$. 

\beq \Phi_{\bs{n}}= \pi^{3/2}2^{-\frac{1}{2}(\sum_{\alpha}n_{\alpha})}
e^{-\frac{1}{2}(\sum_{\alpha}b_{\alpha}x^2)}
\prod_{\alpha}\sqrt{\frac{b_{\alpha}}{n_{\alpha}!}}
H_{n_{\alpha}}(\alpha \sqrt{b_{\alpha}}) \eeq
\beq E_{\bs{n}}=\sum_{\alpha}(n_\alpha + \frac{1}{2})\hbar \omega_{\alpha}\eeq
Here $\alpha=x,y,z$. $H_{n}(\alpha \sqrt{b_{\alpha}})$ is  
a Hermite polynomial of 
degree $n$ and 
\beq \sqrt{b_{\alpha}}=
\sqrt{\frac{m\omega_{\alpha}}{\hbar}} = \frac{1}{a_{\alpha}}
\label{shl} \eeq  where $a_{\alpha}$ is the  
harmonic oscillator length scale corresponding to the axis $\alpha$. 
In this most general case of three distinct trap frequencies, the 
eigenfunctions and energy levels of the  hamiltonian $H_{rot}$
are not  simply related to those corresponding to the non-rotating case.
We shall not discuss this case any further. 
Rather in the following discussion we consider only some 
special cases of interest where such a simple relation exists.

\subsection{Spherical oscillator (SO)}\label{SO}
The hamiltonian is that of an isotropic or spherical oscillator with 
$\omega_x=\omega_y=\omega_z = \omega$. 
The problem is separable in  spherical co-ordinates ($r,\theta, \phi$). 
There
\beq H_{0}=\frac{\hbar^2}{2m}\left(-\frac{1}{r}\frac{d^2}{dr^2}r + 
\frac{\bs{L}^2}{r^2}
+\frac{1}{2}m\omega^2 r^2 \right), \eeq
The eigenfunctions and eigenvalues are described by  
a set of three quantum numbers $\bs{n}=(n_{r},l,l_{z})$
where $n_r \in \bs{N};\; l \in \bs{N};\; -l \leq l_z \leq l$. 
The eigenfunctions are written
as a product of a Laguerre polynomial (radial part) and 
of a Spherical harmonic
(angular part). 
 
\beq \Phi_{\bs{n}}  = b_{\omega}^{3/4}
\frac{1}{\pi^{\frac{1}{4}}}(\sqrt{b_{\omega}} r)^{l} 
e^{-(\frac{r}{b_{\omega}})^2/2}
\sqrt{\frac{2^{n_r+l+2} n_r!}{(2n_r + 2l +1)!!}} L_{n_r}^{l+\frac{1}{2}}
(b_{\omega} r^2) Y_{l,l_z}(\theta, \phi) \label{sphf} \eeq
\beq E_{\bs{n}}=\hbar \omega (2n_{r} + l +\frac{3}{2}) \label{sphe}\eeq
where $L_{n}^{l}$ is a  Laguerre polynomial and $Y_{l,l_z}$ is a 
spherical harmonic. 
These are simultaneous eigenfunctions of the hamiltonian  
and one of the three angular momentum operators. Conventionally, 
this component 
is identified with the $z$-axis namely  
\beq L_{z}\Phi_{\bs{n}}=l_{z}\Phi_{\bs{n}} \eeq
The energy is 
degenerate in $l_{z}$. 
If the spherical trap is rotated about the  
$z$-axis, with a rotational frequency 
$\Omega$ 
\beq H_{rot}=H_{0}-\Omega L_{z} \eeq
The eigenfunctions of $H_{rot}$ are identical to those defined
in (\ref{sphf}) but with the eigenvalues
\beq E^{rot}_{\bs{n}}(\Omega)=\hbar \omega(2n_r +l(1-
\frac{l_{z}\Omega}{l\omega})
+\frac{3}{2}) \label{esr}\eeq
The $l_z$ degeneracy is therefore lifted by the rotation.

\subsection{Two-dimensional isotropic harmonic oscillator (2DIO)}\label{2DIO}
Another case of interest corresponds to $\omega_{x}=\omega_{y}=\omega_{\perp}
\neq \omega_{z}$. This trap geometry is particularly adopted in most of the 
experiments on atomic BEC.
Using cylindrical polar co-ordinates ($r, \theta, z$) 
one finds that the problem separates into a pair of harmonic 
oscillators namely an isotropic oscillator in the $x$-$y$
plane and another one dimensional oscillator along the $z$-axis. The 
eigenfunctions can be written as a product of the 
eigenfunctions of a two-dimensional isotropic oscillator and of a
one dimensional harmonic oscillator. The energy-eigenvalues 
are the sum of the 
corresponding eigenvalues. We also define 
the aspect ratio $\lambda_{R}$
which gives the relative strength of the confinement of the one dimensional
harmonic oscillator in the $z$-direction to the 
confinement of the isotropic oscillator, namely, 
\beq \lambda_{R} = \frac{\omega_{z}}{\omega_{\perp}} \label{AR} \eeq
Since this pair of oscillators with different dimensionalities
are decoupled 
from each other we shall consider only the motion of the isotropic 
harmonic oscillator. 
In $(r, \theta)$ co-ordinates the hamiltonian writes 
as 
\beq H_{\perp}=-\frac{\hbar^2}{2m}\left[ \frac{\partial}{\partial r}(r 
\frac{\partial}{\partial r}) + \frac{L_{z}^2}{r^2} \right]+\frac{1}{2}m
\omega_{\perp}^2 r^2 \eeq
The eigenfunctions and energy eigenvalues can be written in terms
of two quantum numbers $\{ \bs{n}_{\perp} \}=(n, l_z)$ 

\beq
\Phi_{\bs{n_{\perp}}} = \left(
\frac{b_{\omega_\perp}^{l_z+1}}{\pi}\frac{n!(n+l_z)!}{n!^2 l_z!^2}\right)^{1/2}
r^{|l_z|}e^{il_z\theta}
e^{-\frac{b_{\omega_{\perp}}r^2}{2}}
~_1F_1(-n,|l_z|+1;b_{\omega_{\perp}} r^2) \label{2df} \eeq
\beq E_{\bs{n_{\perp}}} = (2n + |l_z| +1)\hbar \omega_{\perp} \label{2de} \eeq
where $b_{\omega_\perp}=\frac{m \omega_\perp}{\hbar}$, 
$n \in \bs{N}$, $l_z \in \bs{Z}$ and 
$_1 F_1(\alpha',\beta';x)$ is a
confluent hypergeometric function. 
Those are eigenstates of the angular momentum operator $L_{z}$
with eigenvalues $l_{z}$. 
If the trap is rotated about the $z$-axis, 
the corresponding hamiltonian is given by 
\beq H_{rot}=H_{\perp} - \Omega L_{z} \eeq
whose  eigenfunctions are identical to those in (\ref{2df}) with the 
eigenvalues 
\beq E^{rot}_{n,l_z}=(2n + |l_{z}|-\frac{\Omega}{\omega_{\perp}}l_z +1)\hbar
\omega_{\perp}
\label{2dre}\eeq
We shall use latter the equivalent expression of the hamiltonian in the 
rotating frame
\beq H_{rot}= \frac{1}{2m}({\bf p} -m {\bf
A}_{\Omega})^2+\frac{1}{2}m(
\omega_{\perp}^2- \Omega^2)r^2 \label{hm3} \eeq
where ${\bf A}_{\Omega}= \bs{\Omega} \times {\bf r} $
and ~$\bs{\Omega} = (0,0,\Omega)$. This is the  
Landau hamiltonian of a particle in a transverse
magnetic field ${\bf B}_{\Omega}~=~2\bs{\Omega}$ 
(written in the symmetric gauge) 
and in a parabolic confinement $\frac{1}{2}m(
\omega_{\perp}^2- \Omega^2)r^2$. 
\subsection{Characteristic frequency of vortex nucleation}\label{subcfsb}
For a spherical oscillator, 
with $\Omega < \omega $, the ground state is $(n, l, l_z)=(1, 0, 0)$ always 
remains the ground state. Similarly the ground state of the 
isotropic oscillator 
such that $\Omega < \omega_{\perp}$ is $(n, l_z)=(0, 0)$.
For a set of non-interacting 
rotating trapped bosons, a vortex is nucleated 
if in the co-rotating frame
a state with higher angular momentum quantum number 
becomes the ground state of the system. If this happens the corresponding 
rotational frequency is the characteristic nucleation frequency of the first 
vortex ($\Omega_{c1}$). According  to this definition,  
in either cases of SO and 2DIO, the 
the first vortex ($l_z=1$) is nucleated when  
the energy of the $l_z=1$
state in the rotating frame is equal to the ground state energy ($l_z=0$). 
Using (\ref{esr}) and (\ref{2dre}) and  
the relation $E^{rot}_{\bs{n}}=E_{\bs{n}} - \Omega l_z$,  
we therefore get  
\beq \Omega_{c1} = \frac{N_0(E_{\bs{n}}(n_{r}=1, l=1, l_z=1)- E_{\bs{n}}
(n_{r}=1,l=0, l_z=0))}{N_0\hbar} = \omega ~~~(SO) \label{cfso}\eeq
\beq \Omega_{c1}= \frac{N_0(E_{\bs{n}}(n=1, l_z=1)-E_{\bs{n}}
(n=1, l_z=0))}{N_0\hbar} = \omega_{\perp} ~~~(2DIO) \label{cfio}\eeq
For either $\Omega = \omega_{\perp}$ or $\Omega=\omega$, the confinement
potential is compensated by the centrifugal force. The trapped condensate 
is therefore unstable at this point.
Therefore, non-interacting bosons in a trap cannot have 
stable vortex solutions at $T=0$.
Since experimentally vortices
are nucleated before this limit is reached and  
at very low temperatures  
($k_{B}T \ll \hbar \omega$) one concludes that a finite amount 
of interactions among bosons leads to the vortex nucleation. 
Thus, in order to study 
vortex nucleation, one needs to include interaction
between bosons. We shall first consider the limit of a weakly 
interacting gas within the framework of the 
Bogoliubov \cite{Bogo47} theory. We shall assume 
repulsive and local interactions among the bosons. 
The next section is devoted to the description of
the Bogoliubov theory.
 
\setcounter{equation}{0}%
\renewcommand{\theequation}{\mbox{\arabic{section}.\arabic{equation}}}%
\section{Bogoliubov Description of a weakly interacting Bose gas}\label{sbd}
The Bogoliubov description of a weakly interacting Bose gas
(with and without confining potential)
has been detailed in  a 
large number of references \cite{ Fetter71, Fetter72, 
BR85, book, book1, Mark96, Fett96, Dalf99, CastinR}.  
We start with the following interacting hamiltonian for a $N$-boson system
\beq H_{mb}=\sum_{i=1}^{N}\left(-\frac{\hbar^2}{2m}\nabla_{i}^2 + 
V_{tr}(\bs{r}_{i})\right)+g\sum_{i,j} \delta(\bs{r_i}-\bs{r_j}) 
\label{MBH}\eeq
where $g=\frac{4\pi\hbar^2 a}{m}$, $m$ is the mass of a boson, $a$ is the 
$s$-wave scattering length which characterizes the 
repulsive delta-function interaction. 
Also
\beq V_{tr}=\frac{1}{2}m(\omega_x^2 x^2 +\omega_y^2 y^2 + \omega_z^2 z^2)
\label{trappot}\eeq  
In a second quantized form, the grand-canonical many-boson hamiltonian is     
\beq \hat{F}= \int d\bs{r} \hpa(\bs{r})\left(-\frac{\hbar^2}{2m}\nabla^2 
+ V_{tr}(\bs{r}) \right)\hp(\bs{r}) + \frac{g}{2}
\int d\bs{r}\hpa(\bs{r})\hpa(\bs{r})\hp(\bs{r})\hp(\bs{r})
-\mu \hat{N} \label{freeen} \eeq
where  
$\hat{N}=\int d\bs{r}\hpa(\bs{r})\hp(\bs{r})$ is the number operator
and $\mu$ is the chemical potential. The bosons field operators satisfy 
\beq [\hp(\bs{r}), \hpa(\bs{r'})] 
=\delta(\bs{r} - \bs{r'})~ ;~ [\hp(\bs{r}), \hp(\bs{r'})] 
= [\hpa (\bs{r}), \hpa(\bs{r}')]=0 \eeq
In general, the field
operator $\hp(\bs{r})$ can be expanded as 
\beq \hp(\bs{r}) = \sum_{\bs{n}}\Phi_{\bs{n}} ({\bf r})  a_{\bs{n}} \eeq  
where $\Phi_{\bs{n}} ({\bf r})$
are single-particle wavefunctions and $a_{\bs{n}}$ are the
corresponding annihilation operators \cite{Fetter71}. They 
obey the commutation rules:
\begin{equation}
\big[ a_{\bs{n}_1}, a_{\bs{n}_2}^{\dagger} \big] =  \delta_{\bs{n}_1, \bs{n}_2}
\; , \;  \big[ a_{\bs{n}_1}, a_{\bs{n}_2} \big]  =  0  \; , \; 
\big[ a_{\bs{n}_1}^{\dagger}, a_{\bs{n}_2}^{\dagger} \big]  =  0 \; .
\label{commut}
\end{equation}
Bose-Einstein  condensation occurs when the number of atoms of a
particular single-particle state becomes very large: $\equiv N_0
\gg 1$ and the ratio $N_0/N$ remains finite in the thermodynamic limit 
$N\to\infty$.
In the limit $N \to \infty$
the states with $N_0$ and $N_0 \pm 1
$ atoms correspond to the same physical configuration. Thus the 
operators  $a_0$ and  $a_0^{\dagger}$ can be considered as $c$-numbers
since their commutator is negligible in comparison to 
$N_0$. Therefore one sets $a_0=a_0^{\dagger}=\sqrt{N_0}$. The above 
definition for the Bose-Einstein condensation holds for 
the non-interacting gas of bosons. 

In presence of the interaction this
definition of Bose-Einstein 
condensation in terms of the macroscopic occupation of a single boson 
eigenstate needs to be generalized. The generalization
is done by requiring that
one of the eigenvalues of the one body density matrix,namely  
the quantity $<\hp^{\dagger}(\bs{r})\hp(\bs{r})>$ is macroscopic
\cite{Pen51}. We shall not discuss this issue here and for details
see chapter $1$ of \cite{book1}.
Thus 
the field operator $\hp(\bs{r})$ can be written as a sum of a c number 
$\P(\bs{r})=\sqrt{N_0}\Phi_{0}$, 
plus a correction $\delta \hp(\bs{r})$,
\beq \hp(\bs{r}) = \P(\bs{r}) + \delta \hp(\bs{r}) \label{dcp} \eeq
Here $\Phi_{0}$ is the macroscopically
occupied single particle state. 
The classical field $\Psi(\bs{r})$ is defined as the condensate 
wavefunction. It is normalized by demanding
\beq \int d\bs{r}|\P(\bs{r})|^2 = N_{0} \label{norm}\eeq 
such that $N_0$ is the total number of particles in the condensate.
Plugging (\ref{dcp}) in the expression for the free energy 
(\ref{freeen}) and neglecting terms higher than quadratic 
in $\delta \hp(\bs{r})$ and its adjoint finally yields, 
\bea 
\hat{F}&=&\int d\bs{r} \P^{\ast}[H_{0} - \mu +\frac{g}{2}|\P(\bs{r})|^2]
\Psi(\bs{r}) \nonumber \\
& & \mbox{+}\int d \bs{r}\left( \P^{\ast}[H_{0} - \mu +g|\P(\bs{r})|^2]
\delta \hp(\bs{r}) + \delta \hpa(\bs{r})[H_{0} - \mu +g|\P(\bs{r})|^2]
\Psi(\bs{r}) \right)\nonumber \\
&+& \frac{1}{2}\int d \bs{r}\left(\begin{array}{cc}
\delta \hpa(\bs{r}) & 
\delta \hp(\bs{r})\end{array}\right)
\left(
\begin{tabular}{cc}
$\displaystyle \mathcal{L}$
& $g \Psi^2(\bs{r})$ \\
$-g  \P^{*2}(\bs{r})$ &
$\displaystyle -[\mathcal{L}]^{\ast}$
\end{tabular}
\right)
\left(\begin{array}{c}
\delta\hp(\bs{r})\\
\delta \hpa(\bs{r}) \end{array}\right) \label{BG1}\eea
where   $\mathcal{L}=-\frac{\hbar^2}{2m}\nabla^2+V_{tr} -\mu +
2g|\P(\bs{r})|^2$.
For later use we shall denote the $2 \times 2$ matrix which appears 
in the above 
expression as ${\cal H}_{B}$ (Bogoliubov hamiltonian).

\subsection{Classical approximation: Gross-Pitaevskii equation}\label{subgp}
The first term in the expression (\ref{BG1}) involves only  
the condensate wave function $\P(\bs{r})$ and 
its minimization gives the stationary 
Gross-Pitaevskii(GP) equation
\cite{Gros61, Pita61}
\beq [H_{0}+g|\P (\bs{r})|^2]\P(\bs{r})=\mu \P(\bs{r}) \label {GPEQ}\eeq
This is also known as the non-linear Schr$\ddot{o}$dinger equation. 
The time dependent GP equation (for example see chapter $7$ \cite{book})
is obtained by replacing 
the $\mu \P(\bs{r})$ by $i\hbar \frac{\partial \Psi(\bs{r},t)}{\partial t}$: 
 \beq i\hbar \frac{\partial \Psi(\bs{r},t)}{\partial t} =
[H_{0}+g|\P (\bs{r})|^2]\P(\bs{r}) \label {TDGPEQ}\eeq
The consistency between two equations requires that under 
stationary conditions $\P(\bs{r}, t)$ must evolve in time 
as $e^{-i\frac{\mu}{\hbar}t}$.
The Gross-Pitaevskii energy functional  
which can be minimized to give the GP equation (\ref{GPEQ})is  
given by 
\beq E_{GP}=\int d\bs{r} 
\left[\frac{\hbar^2}{2m}|\bs{\nabla}\Psi(\bs{r})|^2 +
V_{tr}(\bs{r})|\Psi(\bs{r})|^2
+\frac{g}{2}|\Psi(\bs{r})|^4 \right] \label{gpen1}\eeq
and the associated Lagrangian is 
\beq L_{GP}= \int d\bs{r} \left[\frac{i\hbar}{2}\left
(\P^{\ast}\frac{\partial \P}{\partial t}-
\P\frac{\partial \P^{\ast}}{\partial t}\right)- 
E_{GP}(\P, \P^{\ast}) \right] \label{gpl1} \eeq

The quantity $\Psi(\bs{r})$ is called the condensate wavefunction since 
it represents the coherent motion of all the particles in a condensate. 
It is different from the first quantized many body wavefunction
of the $N$-boson system 
which is a function of the co-ordinates of $N$ bosons and in principle
can be obtained from the diagonalization of the 
hamiltonian of this $N$-boson system (\ref{MBH}). This wavefunction
contains all  correlations present in the many boson system whereas 
the condensate wavefunction only represents the coherent motion of
all condensed particles. We shall not discuss this issue any further
which is detailed in a number of references, for example, (
 chapter $7$ of \cite{BR85}, \cite{Feyn54, ady71}
chapter $5$ of \cite{Noz}).

\subsection{Healing length of the condensate}\label{subhl}
We shall now define the healing length of the condensate from the 
GP equation. For a condensate 
without any confinement potential, the healing length is defined as the length
over which the wave function changes appreciably from its 
unperturbed value under some perturbation.
It describes the size of the vortex core 
and the length over which the density 
grows from zero to its bulk value in the vicinity of a 
boundary \cite{Ginz58}. Without confinement (\ref{GPEQ}) rewrites  
\beq \left[-\frac{\hbar^2}{2m}\nabla^2 + g|\P(\bs{r})|^2  \right]
\P(\bs{r})=\mu \P(\bs{r})
\eeq
The minimum energy configuration corresponds to a uniform condensate density 
given by $|\P(\bs{r})|^2=\rho_0=\frac{\mu}{g}$ since the kinetic energy of
such a configuration vanishes.
If the density is locally perturbed, then far away  from the  
perturbation it approaches this constant value 
$\rho_{0}=\frac{\mu}{g}$ again.
The wavefunction can therefore be 
written as $\P(\bs{r})=\sqrt{\rho_{0}}\psi(\bs{r})$ such that $\psi \rightarrow 1$
when $r \rightarrow \infty $. The GP equation then becomes 
\beq \left[-\frac{\hbar^2}{2m \rho_{0}g} \nabla^2 + |\psi(\bs{r})|^2 \right]
\psi(\bs{r})=\frac{\mu}{\rho_{0} g} \psi(\bs{r})
\eeq
from which we see that the kinetic energy term is appreciable
only if the wavefunction varies beyond the length scale defined by  
\beq \xi =\sqrt{\frac{\hbar^2}{2m\rho_{0}g}}=
\frac{1}{(8\pi\rho_{0}a)^{\frac{1}{2}}} \label{hl} \eeq
$\xi$ is called the healing length or the coherence length of the condensate.

\subsection{Vortex solution of the GP equation $-$ 
Quantization of circulation}\label{V}
The solutions of the GP equation (\ref{GPEQ})
can generally be written as $\P(\bs{r})=|\P(\bs{r})|e^{iS(\bs{r})}$
({\it Madelung transformation}). 
The associated current can be defined (for details, see \S\ref{subhyd} ) as
\beq \bs{j}(\bs{r})=\frac{\hbar}{2im}\left[\P(\bs{r})^{\ast}
\bs{\nabla}\P(\bs{r})- \P(\bs{r})
\bs{\nabla}\P(\bs{r})^{\ast} \right]=
|\P(\bs{r})|^2 \frac{\hbar}{m}\bs{\nabla}
S(\bs{r})=
\rho(\bs{r})\frac{\hbar}{m}\bs{\nabla}
S(\bs{r}) \eeq
The condensate velocity $\bs{v}_{s}(\bs{r})$  
is therefore 
$\frac{\bs{j}(\bs{r})}{\rho(\bs{r})}$
and it is proportional to the gradient of a scalar function. 
The flow of the condensate is therefore irrotational except at the 
points where the phase is singular, namely
\beq \bs{\nabla} \times \bs{v}_{s} = 
\frac{\hbar}{m}\bs{\nabla} \times \bs{\nabla}
S(\bs{r})=0 \label{irotcond}\eeq
This implies that the condensate can be rotated only through the formation 
of vortices such that the condensate density vanishes 
at the vortex cores.
To illustrate this point 
let us assume that the trap is a perfect cylinder such that the 
wavefunction $\Psi$
is an eigenfunction of the operator $L_{z}$. 
In the cylindrical co-ordinate the phase of 
such an eigenfunction can generally be 
written as $S(z,r,\theta)=\kappa\theta$ such that the 
wavefunction is given by 
\beq \P(z,r,\theta)=\P(z,r)e^{i\kappa\theta} \eeq
The single-valuedness of the wavefunction requires 
$\kappa$ to be an integer. This leads to  
\beq \oint _{\Gamma} \bs{v}_{s}. d\bs{r}=\frac{\hbar}{m}\oint_{\Gamma}
\bs{\nabla}S(\bs{r}).d\bs{r} =\oint_{\Gamma}
(\frac{\kappa\hbar}{mr} \hat{\theta}).(dr\hat{r}+r d\theta \hat{\theta})
=\frac{\kappa h}{m} \eeq
Therefore the circulation of the local velocity $\bs{v}_s$ along any 
closed contour $\Gamma$ which includes the point
where the phase is singular is quantized
in  units of $\frac{h}{m}$ \cite{On49, Feyn55}. 

\subsection{Leading order corrections:Bogoliubov equations}\label{BogE}
Coming to the next order correction to the condensate wavefunction, 
one notes that the first integral in  
(\ref{BG1}) is a $c$-number which we called $E_{g}$ 
and the second integral vanishes identically if $\P(\bs{r})$ is 
a solution of the GP equation (\ref{GPEQ}). 
Since the hamiltonian ($H - E_{g}$) (\ref{BG1}) is 
quadratic in $\delta{\hpa}$ and $\delta{\hp}$ it can be written 
in a diagonalized form, namely, 
\beq H - E_{g} = \sum_{\lambda}E_{\lambda}a_{\lambda}^{\dagger}a_{\lambda} 
\label{dbogt} \eeq
through the following Bogoliubov transformations
\beq \delta{\hp}(\bs{r})=\sum_{\lambda}(u_{\lambda}(\bs{r})a_{\lambda}
+v_{\lambda}^{\ast}(\bs{r})a_{\lambda}^{\dagger}) \label{bgt} \eeq
such that $u_{\lambda}$
and $v_{\lambda}$ satisfies the following set of the eigenvalue equations 
\bea \mathcal{L}u_{\lambda}(\bs{r}) + g\P^2(\bs{r})v_{\lambda}(\bs{r})
&=&
E_{\lambda}u_{\lambda}(\bs{r}) \nonumber \\
\mathcal{L}v_{\lambda}(\bs{r}) + g\P^{\ast 2}(\bs{r})u_{\lambda}(\bs{r})&=&
-E_{\lambda}v_{\lambda}(\bs{r}) \label{bogpart}\eea
Unlike ${\cal H}_{B}$ in (\ref{BG1}) which (apart from the c-number $E_{g}$)
defines a set of 
interacting bosons, the hamiltonian    
(\ref{dbogt}) describes non-interacting 
quasiparticles and the state of  
wavefunction $\P(\bs{r})$ which satisfies the
GP equation is the corresponding vacuum. The wavefunction $\P(\bs{r})$
is characterized by its winding number $\kappa$
and for each $\kappa$ have its corresponding Bogoliubov quasiparticles. 
The quasiparticles are bosons and are related 
to the original bosons by the transformation (\ref{bgt}).  
The commutators of 
$a_{\lambda}^{\dagger}, a_{\lambda}$ are same as those for the 
ordinary bosons 
(\ref{commut}). For more details
see \cite{CastinR}.
Since the energy of this 
quasiparticle is defined with respect to the condensate energy $E_g$,
the presence of a Bogoliubov quasiparticle with negative energy and the  
normalization $\langle u_{\lambda}|u_{\lambda}\rangle 
-\langle v_{\lambda}|v_{\lambda}\rangle=+1$
implies an energetic instability for the solution
of the GP equation. Such instability is known as the  
thermodynamic instability. $\Psi$ being a solution of a non-linear
equation there is also another kind of instability which is known
as the dynamical instability. This instability determines whether
there exists time-dependent fluctuations around $\P$ whose
amplitude exponentially
grows over time and destabilizes it. Relation between these two types
of (in)stabilities 
will be discussed in section \S\ref{GPLR}.

\subsection{Bogoliubov theory in a rotating frame}\label{BGROT}
Stable vortex solutions are obtained in rotating trapped condensates
which consist of bosonic atoms (atoms with integer spin). 
Therefore we mention the structure of the Bogoliubov theory in 
a rotating frame which has been studied in \cite{Tomo01, SMS}.  
In the rotating frame the second quantized 
grand-canonical hamiltonian (\ref{freeen}) is 

\beq
 \hat{F}_{rot}  =  \hat{F} -
          \hpa ( \bs {r} ) (\bs{\Omega} \cdot \bs{L})
          \hat{\psi} (\bs{r} ) d\bs{r}  \eeq

After the Bogoliubov decomposition we obtain
\bea
 \hat{F}_{rot} 
   & = & \hat{F} + \int d \bs{r} \;
   \; [\P^{\ast} ( {\bf r} ) (\bs{\Omega} \cdot  \bs{L} ) \P( {\bf r} ) 
     + \delta \hpa( {\bf r} ) (\bs{\Omega} \cdot \bs{L}) \P( {\bf r} ) \\
& + &        \P^{\ast} ( {\bf r} ) (\bs{\Omega} \cdot \bs{L}) 
\delta\hp( {\bf r} ) 
     + \delta \hpa ( {\bf r} ) 
(\bs{\Omega} \cdot \bs{L}) \delta \hp( {\bf r} )
      \;   ]. \eea
where $\hat{F}$ is given by(\ref{BG1}). By keeping only the terms
which involve only the condensate wavefunction, we obtain the 
GP equation in the rotating frame, namely 
\beq [H_{0}+g|\P (\bs{r})|^2 - \bs{\Omega}. \bs{L}]\P(\bs{r})
=\mu \P(\bs{r}) \label {RGPEQ}\eeq
The corresponding Bogoliubov hamiltonian in the rotating frame becomes 
\beq {\cal H}_{B}(rot)=\left(
\begin{tabular}{cc}
$\displaystyle \mathcal{L}- \bs{\Omega} \cdot \bs{L}$
& $g \Psi^2(\bs{r})$ \\
$-g \P^{*2}(\bs{r})$ &
$\displaystyle -[\mathcal{L}-\bs{\Omega} \cdot \bs{L}]^{\ast}$
\end{tabular}
\right) \eeq
The difference between this expression and (\ref{BG1}) is 
the appearance of the term $-\bs{\Omega} \cdot \bs{L}$ in the diagonal
elements. For simplicity let us consider the situation when a 
system is rotated about the $z$-axis so that the extra term is 
$\Omega L_{z}$. Under complex conjugation this operator changes its 
sign making the diagonal elements different in the 
${\cal H}_{B}(rot)$ whereas they
are identical in ${\cal H}_{B}$ apart from an overall $-$ sign. 

\subsection{Characteristic frequency of vortex nucleation}\label{gpcritf}
The solution of the 
GP equation in the rotating frame is used to determine the 
characteristic frequency of vortex nucleation. 
The principle is the following. Let us again consider the case 
of a cylindrical symmetry (\S\ref{V}).
One determines the energies (\ref{gpen1}) of a condensate with and without
a vortex  
by respectively setting $\kappa$ equal to some non-zero integer and $\kappa=0$
in the expression for $\Psi$. These energies are respectively denoted as
$E_{v}(\kappa)$ and $E_0$. 
For a static trap $E_v(\kappa)$ is always greater than 
$E_0$. In a rotating trap 
at a given value of $\Omega = \Omega_{c \kappa}$, $E_v(\kappa)$
becomes the ground state energy of the hamiltonian in (\ref{RGPEQ}). 
The corresponding angular momentum is $\kappa \hbar N_0$ where $N_0$
is the number of atoms in the condensate. Therefore the characteristic
rotational frequency for the nucleation of a vortex of winding 
number $\kappa$ is given by \cite{sd96}
\beq  \Omega_{c \kappa}=\frac{E_v(\kappa) - E_0}
{N_0\kappa \hbar} \label{critfreq} \eeq
The ground state energy of the hamiltonian is also the minima
of the thermodynamic free energy. Thus
this frequency is also called the thermodynamic frequency
of vortex nucleation since it is determined by minimizing the thermodynamic
free energy.
Using this formula
Dalfovo {\it et. al.} \cite{sd96, dps96} 
evaluated  numerically the critical frequency of vortex nucleation. 
Their treatment of the full GP equation (\ref{GPEQ})
includes the effect 
of the confinement as well as that of  
the interaction. To understand the effect
of each of these terms we start with the case where 
both the interaction 
and the confinement are neglected and then subsequently include 
their effect in steps.

\begin{itemize}
\item {\bf Non-interacting case} 
\end{itemize}
To describe the non-interacting case \cite{Noz}
we choose again the geometry of 
an infinitely long cylinder. There the superfluid flow is irrotational
(\ref{irotcond}), if the magnitude of 
velocity varies as 
\beq \bs{v}_{s}=\frac{c}{r} \hat{\theta}\label{sfvel1} \eeq
where $c$ is a constant. 
The above equation is valid for large distances. For small distances
{\it i. e.} 
when $r$ becomes comparable to the healing length $\xi$ (\ref{hl}), 
$v_{s}$ varies
rapidly and this definition is no longer valid. 
The quantization of circulation (\S\ref{V})
determines the constant c
to be  
\beq v_{s}=\frac{\kappa\hbar}{mr} \label{sfvel2} \eeq
If the radius of the cylinder is $b$, the kinetic 
energy per unit length along the cylinder axis is given by 
\beq E_v(\kappa)
=\int_{0}^{b}2\pi r dr \frac{\rho m}{2}v_{s}^2=\frac{\pi \rho}
{m}\kappa^2 \hbar^2
\int_{0}^{b}\frac{dr}{r}=
\frac{\pi \rho}{m}\kappa^2 \hbar^2 \log\frac{b}{\xi} \label{ev1}\eeq
The integral is divergent. 
It is regulated by using  
a finite cut-off $\xi$ at the lower limit of $r$.
This is because in presence of vortex the density $\rho$ vanishes at 
the origin at a rate faster than $r$. 
The total angular momentum density
per unit length of the cylinder  is 
$\rho \pi b^2 \kappa \hbar$. Setting 
$\kappa=1$  and using (\ref{critfreq})
one gets the  thermodynamic frequency for the nucleation of 
the first vortex  
\cite{Pethick96}
\beq \Omega_{c1}=\frac{\hbar}{mb^2}\log\frac{b}{\xi} \eeq 

\begin{itemize}
\item {\bf Uniform condensate (unconfined) with interaction}
\end{itemize}
Now let us consider an uniform 
gas of interacting bosons without confining potential $V_{tr}$. 
The geometry is again a cylinder with radius $b$. We evaluate the 
energy of the condensate in the presence of a vortex 
relative to the vortex free condensate with 
the same number of particles. 
The condensate wavefunction  is 
given in (\ref{GPANSATZ})
and because of the symmetry along $z$-axis and very low temperature
it is justified to consider 
both the condensate 
and the vortex state correspond to $k_{z}=0$.
The extra energy per unit 
length of the vortex with $\kappa=1$ (\ref{gpen1})
(chapter 9 of \cite{book}).
\bea E_{v} &=& \int_{0}^{b} 2 \pi r dr 
\left[\frac{\hbar^2}{2m}|\bs{\nabla}\Psi(\bs{r})|^2
+g|\Psi(\bs{r})|^4 \right ] -~E_{0}\nonumber \\
&=& \int_{0}^{b} 2 \pi r dr \left[\frac{\hbar^2}{2m}
\left( \left[\frac{d\Psi(r)}{dr} \right]^2 + \frac{\Psi(r)^2}{r^2}
\right) +g|\Psi(\bs{r})|^4 \right] -~E_{0}\eea
Here $E_{0}$ is the energy of the bare condensate. 
This was evaluated first by  
Ginzburg and Pitaevskii 
\cite{Ginz58} by using the  
solution of the GP equation (\ref{GPEQ}) 
and the result is (for $\kappa =1$) 
\beq E_{v}=\pi \rho \frac{\hbar^2}{m}\log\left( 1.464 \frac{b}
{\xi}\right) \label{GINZPITA} \eeq
where $\rho$ corresponds to the vortex free uniform density of the condensate.
In comparison to the expression (\ref{ev1}) here we can see an
extra factor $1.464$ which accounts for the interaction among bosons.  
The corresponding characteristic frequency of the 
first vortex nucleation is 
\beq \Omega_{c1}=\frac{\hbar}{mb^2}\log\frac{1.464b}{\xi} \eeq
The calculation is detailed in the chapter $9$ of 
\cite{book} and the chapter $5$ of \cite{book1}. 

\begin{itemize}
\item{\bf $\Omega_{c1}$ from the GP equation}
\end{itemize}
The method adopted in \cite{sd96, dps96} by Dalfovo {\it et. al.}
to evaluate the characteristic frequency
for the vortex nucleation for a trapped non-uniform
condensate from the GP equation is very similar to the one 
described for the uniform interacting condensate. They use a geometry
with an axial symmetry (\S\ref{2DIO}). 
For a given number
of particles in the condensate ($N_{0}$)
they solve numerically 
the GP equation to obtain the condensate wavefunction $\P(\bs{r})$
in the presence of a vortex with winding number $\kappa$
and then use the solution to evaluate the 
energy functional (\ref{gpen1}). 
The characteristic
frequency is obtained from the relation (\ref{critfreq}). 
In their work \cite{sd96} $\frac{\Omega_{c1}}{\omega_{\perp}}$ 
is plotted as a function of $N_{0}$.
For $N_{0}=1$ 
the result gives back $\Omega_{c1}=\omega_{\perp}$,
which is the result for a set of non-interacting bosons 
and can be obtained from the 
solutions of the linear (single boson) Schr$\ddot{o}$dinger equation (\ref{cfio}). For $N_{0} > 1$, because of the finite interaction present among bosons, 
the ratio $\frac{\Omega_{c1}}{\omega_{\perp}}$ is
smaller than $1$
making it possible for trapped  
condensate to have a vortex. A numerical investigation of the 
characteristic nucleation frequency and angular momentum 
for a multiple vortex configuration was subsequently done
by Butts and Rokshar \cite{BR99}.

The reason for the reduction 
of $\Omega_{c1}$ in the presence of interactions 
can be understood as follows
\cite{BR99}. In an axially symmetric geometry and 
using the basis of the single 
particle states having definite angular momentum (\ref{2df}) 
(for simplicity we keep the quantum numbers $n$ and $n_{z}$ fixed)
the condensate wavefunction can be expanded as 
\beq \P(\bs{r})=\sum_{l_{z}=0}^{\infty}C_{l_z}\phi_{l_z} \label{expcon}\eeq
The coefficient $C_{l_{z}}$ has to be fixed variationally 
by extremizing 
the following GP energy functional (\ref{gpen1}) in the rotating frame 
\beq E_{GP}(\Omega)=\int d\bs{r} 
\left[\frac{\hbar^2}{2m}|\bs{\nabla}\Psi(\bs{r})|^2 +
V_{tr}(\bs{r})|\Psi(\bs{r})|^2
+\frac{g}{2}|\Psi(\bs{r})|^4 -\P(\bs{r})^{\ast}
\Omega L_{z}\P(\bs{r}) \right] \label{rgpen1}\eeq
To obtain the energy of a vortex state with winding number 
$\kappa$ this extremization 
has to be done with the constraint for the vorticity of 
the condensate wavefunction, namely
\beq \sum_{l_{z}}|C_{l_z}|^2 l_{z}=\kappa \eeq
For a state with no vortex the only way to satisfy  this 
constraint is to make $C_{l_z}=0$ for $l_z \neq 0$ in (\ref{expcon}).
For $\kappa \neq 0$, however it is possible to make $C_{l_z} \neq 0$
for more than one value of $l_z$ provided it is energetically favored.
For non-interacting bosons (\ref{2de}), such an option is 
energetically ruled out upto $\Omega = \omega_{\perp}$. But 
for a set of bosons interacting via the contact interaction (\ref{MBH})
occupying more than one angular momentum orbits reduces the interaction
energy greatly. Therefore a vortex state becomes preferable as soon 
as the reduction in the interaction energy due to such distribution
offsets the gain in the single particle energy (\ref{2de}) due to the 
occupation of the higher $l_z$ states. As a result a 
vortex can be nucleated for 
$\Omega < \omega_{\perp}$. Later we shall show
 how an analytical expression for  
$\Omega_{c1}$ is obtained in
the strongly  interacting limit that agrees very well with the result
of \cite{sd96, dps96} for large $N_{0}$.  
\setcounter{equation}{0}%
\renewcommand{\theequation}{\mbox{\arabic{section}.\arabic{equation}}}
\section{Gross-Pitaevskii equation and its linear stability analysis}
In the section \S\ref{BogE} we have discussed the thermodynamic stability
of the condensate wavefunction $\P$ and mentioned that it is related to 
a dynamic stability. To check the dynamic 
stability of a solution  a linear stability 
analysis of a steady state solution of the time-dependent GP equation
(\ref{TDGPEQ}) has to be performed.   
\cite{Mark96, Dodd97, Sinh97, FR98, SF98, CastinR}. Such an analysis
also gives the linear response of the condensate to an external 
perturbation. Particularly the 
dynamical instability associated
with a particular type of steady state solution leads to the vortex
nucleation as we shall see. In the following section 
we describe the linear stability analysis. 

\subsection{Linear response analysis of the Gross-Pitaevskii equation}\label{GPLR}  
Let us consider the 
effect of adding a weak, sinusoidal perturbation to the 
trap potential. Following Edwards {\it et.al.}
\cite{Mark96} one can write the time-dependent GP equation (\ref{TDGPEQ}) 
under such weak perturbation
\beq i\hbar \frac{\partial \P(\bs{r},t)}{\partial t}= 
\left [ (-\frac{\hbar^2}{2m}\nabla^2 + V_{tr}(\bs{r})
+{\it f}_{+}(\bs{r})\exp(-i \omega_{p}t)+{\it f}_{-}
(\bs{r})\exp(+i \omega_{p}t) )\right ] \P(\bs{r},t) \label{lres1} \eeq
where ${\it f}_{\pm}$ and $\omega_{p}$ 
are respectively the amplitudes and the frequency 
of the sinusoidal perturbation. 
Since it is weak one assumes that 
the deviation from the solution of
the unperturbed GP equation is also going to be small
and can be written under the form 
\beq \Psi(\bs{r}, t)=\exp(-i\mu t)\left [ \P(\bs{r}) + 
\delta \Psi(\bs{r},t) \right ] \label{lres2}\eeq
with 
$ \delta \Psi = u(\bs{r})\exp(-i\omega_{p}t)+
v(\bs{r})^{\ast}\exp(i \omega_{p}t) $. 

After inserting Eq. (\ref{lres2}) into Eq. (\ref{lres1}) 
and keeping terms which are linear in 
$u(\bs{r}), v(\bs{r}), {\it f}_{\pm}(\bs{r})$ one gets 
\beq {\cal H}_{B}\left(\begin{array}{c}
u(\bs{r})\\
v(\bs{r})\end{array}\right)=
\left(\begin{array}{c}
\P(\bs{r}){\it f}_{+}(\bs{r})\\
\P^{\ast}(\bs{r}){\it f}^{\ast}_{-}(\bs{r}) \end{array}\right ) \eeq
The Bogoliubov hamiltonian ${\cal H}_{B}$ 
therefore appears in the linear stability analysis. 
The equations which determine the time evolution of the small fluctuations 
$u$ and $v$ are identical to the  equations (\ref{bogpart}) 
(after ${\it f}_{\pm}(\bs{r})$ is set to zero and 
after changing $p$ into $\lambda$). 
The linearized evolution of the non-condensed part
of the Bose-field operator $\delta \hp(\bs{r})$  which has been 
discussed earlier (section \S\ref{BogE})
is therefore formally equivalent 
to the linearized response of the condensate to a classical perturbation. 
Using Eq. (\ref{TDGPEQ}). 
\beq i\hbar \frac{\partial}{\partial t}
\left(\begin{array}{c}
u_{\lambda}(\bs{r})\\
v_{\lambda}(\bs{r})\end{array}\right)=
{\cal H}_{B}
\left(\begin{array}{c}
u_{\lambda}(\bs{r})\\
v_{\lambda}(\bs{r})\end{array}\right) = \hbar\omega_{\lambda} 
\left(\begin{array}{c}
u_{\lambda}(\bs{r})\\
v_{\lambda}(\bs{r})\end{array}\right) \eeq
The time evolution 
of these modes is given by $e^{-i\omega_{\lambda} t}$. This factor 
remains bounded in time provided that the imaginary part of 
$\omega_{\lambda}$ is negative. This leads  to the dynamic stability 
condition that for all $\omega_{\lambda}$ $Im(\omega_{\lambda}) \le 0$. 
Since the condensate wavefunction in a steady state is obtained by finding
the local minima of the GP energy functional it can be shown that the dynamic
stability criterion is more stringent and requires \cite{Cast99} 
$Im(\omega_{\lambda}) = 0$. 
Using the normalization condition for Bogoliubov wavevectors $u_{\lambda}$
and $v_{\lambda}$
one can  write the condensate response
to an arbitrary small perturbation $f(\bs{r})$ of frequency $\omega_{p}$
as \cite{Mark96}
\beq 
\left(\begin{array}{c}
u(\bs{r})\\
v(\bs{r})\end{array}\right)=
\sum_{\lambda}\frac{g_{\lambda}}{\hbar(\omega_{\lambda} - \omega_{p})}
\left(\begin{array}{c}
u_{\lambda}(\bs{r})\\
v_{\lambda}(\bs{r})\end{array}\right) \label{lrf} \eeq
where 
\beq g_{\lambda}=\int d \bs{r} f(\bs{r})
\left[\Psi(\bs{r})u^{\ast}_{\lambda}(\bs{r})
+\Psi^{\ast}(\bs{r})v^{\ast}_{\lambda}(\bs{r}) \right]\eeq
In the next section we shall describe how 
these normal modes of the Bogoliubov hamiltonian 
are calculated for a set of interacting bosons 
without any confinement which gives 
the usual Bogoliubov dispersion law. This is a standard
text-book material and is given in many references. For example
see \cite{book, book1, CastinR}.

\subsection{Condensate in a box}\label{Bogobox}
The atoms are trapped in a cubic box of size $L$, and we assume
periodic boundary conditions. Solutions of the 
stationary Gross-Pitaevskii equation (\ref{GPEQ}) can be written as
plane waves
with vanishing momentum,
\beq \Phi_0(\bs{r}\,) = \frac{1}{L^{3/2}}\exp(i\bs{k}\cdot\bs{r})|_{\bs{k}=0}
\eeq 
with the chemical potential
$\mu = gN_0 |\phi_0|^2 =\rho_0 g $
and the density of the condensate atoms $\rho_0= N_0/L^3$. 
In the absence of 
any confinement the Bogoliubov hamiltonian  ${\cal H}_{B}$
is translational invariant. Therefore one seeks
its eigenvectors in the form of plane waves.
\beq
\left(\begin{tabular}{c}
$u_{\bs{k}}(\bs{r}\,)$ \\ \\ $v_{\bs{k}}(\bs{r}\,)$
\end{tabular} \right)=
\frac{e^{i\bs{k}\cdot\bs{r}}}{L^{3/2}}
\left(\begin{tabular}{c}
$U_{\bs{k}}$ \\ \\ $V_{\bs{k}}$
\end{tabular} \right).
\label{eq:planew}
\eeq
${\cal H}_{B}$ can be written in a block-diagonal form
with each block ($2 \times 2$ matrix)
corresponding to a given wave vector $\bs{k}$. 
\beq
{\cal H}_{B}[\bs{k}]=
\left(\begin{array}{cc}
\left[\displaystyle\frac{\hbar^2k^2}{2m}+\rho_0 g\right] & \rho_0 g \\
-\rho_0 g & -\left[\displaystyle\frac{\hbar^2k^2}{2m} 
+\rho_0 g\right]\end{array}\right).
\label{2b}
\eeq
This matrix can be diagonalized, 
giving one eigenvector with 
the positive eigenvalue $\epsilon_{\bs{k}}$ and one eigenvector 
with negative eigenvalue $-\epsilon_{\bs{k}}$ where 
\beq
\epsilon_{\bs{k}}=\left[\frac{\hbar^2k^2}{2m}\left(
\frac{\hbar^2 k^2}{2m}+2\rho_0 g\right) \right]^{1/2}
\label{bspectrum}
\eeq
The spectrum (\ref{bspectrum}) was first derived by 
Bogoliubov. Here only the case for the repulsive bosons ($g > 0$)
is considered. For attractive bosons ($g < 0$)
the spectrum exhibits a different behaviour. 
The Bogoliubov spectrum of repulsively interacting
Bose gas 
strongly differs from those of a free Bose gas particularly 
for small $k$. Here the spectrum is linear
and neglecting the quadratic term in  
(\ref{bspectrum}) one gets 
\beq
\epsilon_{\bs{k}} \simeq \hbar k \sqrt{\frac{\rho_0g}{m}}.
\eeq
This corresponds to a spectrum of  
sound-waves propagating 
with a sound velocity $c_s$ given by
\beq
c_s = \frac{d\omega_{\vec k}}{dk}=\frac{1}{\hbar}
\frac{d\epsilon_{\vec k}}{dk} = \sqrt{\frac{\rho_0g}{m}}
\label{soundvel}\eeq
It is this linear part of the spectrum which differentiates the 
behaviour of an interacting Bose gas from that of a free Bose gas. 
For large $k$ the spectrum is quadratic like that of a free Bose gas. 

\subsection{Thermodynamic instability of a vortex solution}
Dodd {\it et. al.} \cite{Dodd97}
carried out the linear stability analysis of the condensate 
with a vortex (with $\kappa$= 1) 
and find out that the corresponding Bogoliubov 
spectrum differs significantly in terms of the 
excitation frequency from Bogoliubov spectrum for the condensate 
without a vortex. Apart from providing 
the clue for the spectroscopic signature of the presence of a vortex
their calculation also revealed the presence of a mode with negative
energy and positive normalization in the spectrum which suggests
that the vortex solution is dynamically unstable. 
Rokshar \cite{Rokh97} (see also \cite{ALF98}) 
explained this instability 
by pointing out that a Bogoliubov
quasiparticle feels an effective potential which is 
approximately of the form (see the diagonal term of ${\cal H}_{B}$) 
\beq V_{eff}=V_{tr} +2g|\Psi(\bs{r})|^2 - \mu \eeq 
where $\Psi(\bs{r})$ is the condensate wavefunction with 
a vortex. Since the condensate density vanishes at the core of 
a vortex and the confinement potential also reaches it minimum there, 
$V_{eff}$ can be negative in the core (relative to the 
chemical potential) forming a bound core 
state. 
The quasiparticles are also bosons. So the
occupation of this  bound state 
can be macroscopic and 
all the particles which form a vortex
can be transferred to this core state. As a result the vortex collapse. 
For non-interacting bosons this 
can be easily verified since the core state is nothing but the ground state
which has always a lower energy than the vortex state for a confined 
system (see the discussion in sec. \S\ref{subcfsb}). 
This has been verified by Dodd {\it et. al.}\cite{Dodd97}
by showing that in the limit of zero interaction, the energy 
corresponding to this mode
approaches the value $-\hbar \omega_{\perp}$.
The minus sign occurs since the energy is measured with respect
to the vortex energy. Therefore in a static trap
a vortex state is always thermodynamically as well as dynamically unstable. 

Before  pointing out how this instability can be removed we 
briefly describe how Bogoliubov
modes are calculated for a trapped condensate with or without a vortex. 
This has been done  by several 
groups \cite{Mark96, Dodd97, Sinh97, SF98, Tomo99, SMS} 
using various schemes to
study the stability properties of the condensate and its vortices. 

Let us assume that the atomic cloud is  
trapped in a cylindrical trap. This happens when the aspect ratio (\ref{AR})
$\lambda_{R} \ll 1$ such that  the confinement potential is given by
\beq
 V_{tr}( {\bf r} ) = \frac{1}{2} m \omega_{\perp}^2 (r^2 + \lambda_{R}^2 z^2)
\approx \frac{1}{2} m \omega_{\perp}^2 r^2 \eeq
and periodic boundary conditions are imposed along the $z$ direction.
The condensate wave function is written in the form:
\beq
\P(\bs{r})=\P( r,\theta,z ) =  \P( r ) e^{i \kappa\theta}e^{i k_{z} z}
\label{GPANSATZ}\eeq
The winding number $\kappa$ (\S\ref{V})
is finite when a vortex is present and vanishes
in the absence of a vortex. Since the trap is axisymmetric 
\bea
 u_{\lambda} ( {\bf r} ) & = & u_{\lambda} ( r )
            e^{i k_z z} e^{i ( l_z + \kappa ) \theta} \\
 v_{\lambda} ( {\bf r} ) & = & v_{\lambda} ( r )
            e^{i k_z z} e^{i ( l_z - \kappa ) \theta}.
\label{normmode} \eea

The quantum numbers 
 ${\bf n}= ( n , l_z , k_z ) $ may take the following values:
 $ n  \in \bs{N}$;
 $l_z \in \bs{Z} $; and
  $  k_z= q \frac{2\pi}{L}, q \in \bs{Z}$, 
where $L$ is the length of the cylinder.

First we solve the GP equation
with the ansatz (\ref{GPANSATZ}) and then we diagonalize the Bogoliubov
matrix ${\cal H}_{B}$ with that solution after expanding 
$u_{\lambda}(r)$ and $v_{\lambda}(r)$ in a suitable basis. This has 
been detailed in a number of references (for example see 
\cite{Mark96, Tomo99}). Since the normal modes are eigenfunctions
of the angular momentum operator $L_{z}$, according to (\ref{normmode})
if $u_{\lambda}(\bs{r})$ is an eigenfunction of $L_z$ with a particular
eigenvalue $l_z^{u}$ (in the unit of $\hbar$), then $v_{\lambda}(\bs{r})$
will be also an eigenfunction of $L_{z}$ 
with eigenvalue $l_{z}^{u} - 2 \kappa$. 

We shall now describe how we can stabilize 
a vortex solution. Following Fetter and Svidzinsky \cite{fsr01}
we denote the anomalous mode frequency as $\omega_{a}$.
One way of stabilizing this mode is to rotate the condensate
\cite{fsr01, Tomo01}.
Let us consider an 
axisymmetric condensate in rotational equilibrium at an angular velocity $
\Omega$ around the $z$-axis. Since in the rotating frame, the hamiltonian
is $H-\Omega L_z$, and the Bogoliubov amplitudes have frequencies $
\omega_{\lambda}(\Omega) 
= \omega_{\lambda} -l_{z}\Omega$, where $\omega_{\lambda}$ is the frequency
in the non-rotating frame and $l_{z}$ is the angular momentum quantum
number. It has been found in \cite{Dodd97} that for the anomalous mode 
$l_{z}=-1$. The resulting frequency in
the rotating frame is
\beq
\omega _a(\Omega )=\omega _a+\Omega ,
\eeq
Since
$\omega _a$ is negative, the anomalous frequency in a rotating frame
increases linearly towards zero with increasing $\Omega $ and particularly $
\omega _a(\Omega )$ vanishes at a characteristic rotation frequency

\beq
\Omega_{c1} = -\omega_a = |\omega_a|
\end{equation}
which implies that for $\Omega \ge \Omega_{c1}$ the
singly quantized vortex becomes locally stable.
Linn and Fetter \cite{LF99} carried out an explicit perturbative analysis 
to confirm this directly from the GP equation.
One can also stabilize the vortex solution by increasing temperature 
and also by putting disorder which will pin the vortices\cite{Tomo01}.
The dependence of the vortex stability on rotation and temperature
are combined to give a phase diagram for the vortices
\cite{Tomo01,S99}.

\setcounter{equation}{0}%
\renewcommand{\theequation}{\mbox{\arabic{section}.\arabic{equation}}}
\section{Approximation schemes to study the behaviour of the condensate}
The Bogoliubov theory gives a very accurate description of the 
trapped BEC near the zero temperature, in terms of
the condensate wavefunction $\P(\bs{r})$ 
and a set of collective excitations.
It also tells whether a state represented by the condensate wavefunction
is stable or not. Present experiments are done under conditions where
further simplifications of this theory is possible. This enables 
us to interpret different experimentally measurable
properties in terms of the known system parameters, like the number 
of atoms in the condensate $N_0$, the interaction strength ($a$) etc. We shall
now discuss these approximation schemes in some detail.

For further approximation one notes that
depending on the 
parameters such as the number of atoms ($N_0$) present, 
the trap frequency etc.  the role
of interaction in determining the properties of an atomic condensate
can be different \cite{Pethick96, sd96}. 
For the trapped condensate this was first pointed out
by Baym and Pethick \cite{Pethick96} and subsequently verified 
numerically by Dalfovo and Stringari \cite{sd96}.
We start with the following Gaussian variational ansatz 
for an isotropic confinement (the present 
derivation follows a review by Castin \cite{CastinR})
\beq
\P(\bs{r}\,) = {1\over \pi^{3/4}R^{3/2}}e^{-r^2/2R^2}
\eeq
where the spatial width $R$ is the only variational parameter.
After substitution in (\ref{gpen1})  this gives 
the following mean energy per particle 
\beq
\overline{e}\equiv {H[\P,\P^*]\over N_0} =
\frac{3\hbar^2}{4mR^2} +\frac{3}{4} m\omega^2R^2 +
{\hbar^2\over m} \frac{N_0a}{R^3}\frac{1}{\sqrt{2\pi}} \label{tfenergy1}
\eeq
Setting the unit of length to 
$\frac{1}{\sqrt{b_{\omega}}}=(\hbar/m\omega)^{1/2}$
and that of energy to $\hbar\omega$ 
one obtains:
\beq
\overline{e} = {3\over 4}\left[\frac{1}{R^2}+ R^2\right] +
\frac{\chi}{2R^3}
\label{tfape}\eeq
with 
\beq
\chi= \sqrt{2\over\pi}{N_0a\over \sqrt{\hbar/m\omega}}.
\label{eq:def_chi}
\eeq
The kinetic energy term
scales like $1/R^2$, whereas 
the energy due to the confinement potential scales as $R^2$.
Since $N_0/R^3$ is the density of atoms in the condensate, 
parameter $\chi$ ($\propto N_0a$)
measures the effect of the interactions on 
the condensate density. 
The case $\chi\ll 1$ corresponds to the weakly interacting regime, 
close to the ideal Bose gas limit
$\chi=0$; the case $\chi\gg 1$ corresponds
to the strongly interacting regime.

\subsection{The Thomas-Fermi approximation for an isotropic confinement}
\label{subtfa}
When $\chi \gg 1$ the condensate can be well described by the 
Thomas-Fermi (TF) approximation. Since $b_{\omega}=1$, 
$\chi \gg 1 \equiv  {N_{0}a} \gg 1 $.
If $N_0$ is now increased, 
to minimize the energy (\ref{tfape}) $R$ 
will take a larger value. The size of the 
condensate $R$ becomes larger and the kinetic 
energy term becomes less important compared to other terms since 
\bea
R & & \gg 1/\sqrt{b_{\omega}} \nonumber \\
\Rightarrow \frac{E_{\mbox{\scriptsize kin}}}{E_{\mbox{\scriptsize harm}}} & & \simeq
\frac{\frac{\hbar^2}{mR^2}}{m\omega^2 R^2} \simeq \left(
\frac{\hbar}{m\omega R^2}\right)^2 \ll 1
\label{TFA}\eea
This result should be contrasted against 
that of the noninteracting case for which the ratio 
is $1$.
Therefore the  kinetic energy term 
in the GP equation (\ref{GPEQ}) can be neglected in this limit.
This is known as the Thomas-Fermi (TF) approximation.
We emphasize that TF regime is called strongly interacting, not
due to a larger value of the $s$-wave scattering length $a$, but
for the dominant role played by the interaction energy 
in comparison to the kinetic energy.  
The Thomas-Fermi radius $R$ which  gives the size 
of the system in this limit, can then be obtained by equating the 
energy due to the confinement potential and the 
repulsive interaction energy. The TF approximation leads to   
\beq
\mu_{TF}\P(\bs{r}\,) 
\simeq V_{tr}(\bs{r}\,)\P(\bs{r}) + g |\P(\bs{r}\,)|^2
\P(\bs{r}\,).
\label{chemtf}\eeq

\beq \Rightarrow
\P(\bs{r}\,) = \left(\frac{\mu_{TF}-V_{tr}(\bs{r}\,)}{g}\right)^{1/2}
\label{tf}
\eeq
The chemical 
potential can now be expressed in terms of the 
other system parameters from the normalization condition
$\int d \bs{r} |\P(\bs{r})|^2 = N_{0}$ where the domain of 
integration is limited by the Thomas-Fermi radius $R$ which is then 
determined from the boundary condition 
$\P(\bs{r})=0$ for $r \geq R$ which yields
\beq \mu_{TF} =\frac{1}{2}m\omega^2 R^2 \label{tfbc}\eeq
Therefore the TF approximation determines the chemical potential $\mu_{TF}$
and the size of the cloud $R$ in terms of $m$, $\omega$, $g$, $N_0$
which are all known parameters. 
The ground state condensate 
wavefunction $\Psi(\bs{r})$ is consequently determined as a 
real valued quantity which vanishes sharply at the boundary $r=R$.
However (see \S\ref{V} ) $\P(\bs{r})$ is generally a complex
valued function which can be written as 
$\Psi(\bs{r})=\sqrt{\rho(\bs{r})}e^{iS(\bs{r})}$.

\subsection{Non-isotropic traps} 
For a non-isotropic harmonic confining potential (\S\ref{ANSO})  
the boundary condition (\ref{tfbc})  
yields the 
shape of the condensate
as being that of an ellipsoid of radius $R_\alpha$
along the axis $\alpha$
\beq
R_{\alpha}^2= \frac{2 \mu_{TF}}{m \omega_\alpha}
\label{radius}
\eeq
where $\alpha=x, y, z$.
Using the same method as in the case of an isotropic confinement
the condensate density and the chemical potential can be determined 
in terms of the system parameters such as $N_{0}$, $a$ and $\omega_{\alpha}$, 
giving respectively 
\beq
\rho(r)=|\Psi_{TF}(\bs{r})|^2 = \frac{1}{g}\,
\left[\,\mu_{TF} - V_{tr}(\bs{r})\,\right]\Theta\left[\,\mu_{TF} - V_{tr}(%
\bs{r})\,\right]= 
\rho(0)\left(1-\sum_{\alpha}\frac{r_{\alpha}^2}{R_{\alpha}^2}\right)
\Theta\left(1-\sum_{\alpha}\frac{r_{\alpha}^2}{R_{\alpha}^2}\right),
\label{TFDP} \eeq
\beq
\mu_{TF}=\frac{1}{2} \hbar\bar{\omega}
\left[15\frac{N_0a}{\left(\hbar/m\bar{\omega}\right)^{1/2}}\right]^{2/5}
=\frac{1}{2}\hbar \bar{\omega}\frac{R_{TF}^2}{\bar{a}^2}
\eeq
where $\bar{\omega}=(\omega_x \omega_y \omega_z)^{1/3}$, 
$\bar{a}=\frac{1}{\sqrt{b_{\bar{\omega}}}}= 
\sqrt{\frac{\hbar}{m\bar{\omega}}}$, and 
the mean TF radius 
$R_{TF}=\displaystyle\prod_{\alpha}(R_{\alpha}) ^{1/3}$. 
In the TF regime, since $\chi\gg 1$, the chemical 
potential $\mu_{TF}$ satisfies
\beq
\mu_{TF} \gg \hbar \bar{\omega},
\eeq 
At the center of the TF ellipsoid ($x,y,z=0$) 
the density (\ref{TFDP})is
maximum and it is given as
\beq \rho(0)= \frac{\mu_{TF}}{g}  \label{tfmu1}\eeq
Unlike for an unbounded and uniform condensate, 
in a trapped non-uniform condensate
the healing length (\ref{hl}) can only be defined in terms of a density
at a reference point. Conventionally it is chosen as $\rho(0)$, the TF
density at the center of the condensate. 
Therefore we set $\xi~=\xi_0=~[8\pi \rho(0)\,a]^{-1/2}$. This leads to the
following relation between the TF chemical potential (\ref{tfmu1}) and the
healing length
\beq  \frac{\hbar^2}{2m \xi_0^2} = g\rho(0)=\mu_{TF} \label{hlmu}
\eeq
Once the healing length is defined, the relation among different length scales 
in the TF regime can be determined from the relation 
\beq
\xi \,R_{TF} = \bar{a}^2 \eeq
Since in the TF regime $\frac{R_{TF}}{\bar{a}} \gg 1$, one has 
\beq \frac{\xi}{\bar{a}%
}=\frac{\bar{a}}{R_{TF}}\ll 1.
\label{tflength} \eeq
$\xi_0$ (\ref{hl}) gives the size of a vortex core. This can be 
checked from the GP equation. 
A small 
healing length $\xi_0$ characterizes a small
vortex core. In contrast, the healing length (and the 
vortex-core radius) in the
ideal Bose gas limit are 
comparable with $\bar{a}$ and hence with the size of the
condensate.

The TF approximation (\ref{TFA}) breaks down at the 
boundary of the TF region where the density 
$|\P(\bs{r})|^2$ drops to zero (\ref{tf}-\ref{tfbc}). 
The realistic wavefunction which can be obtained from the solution
of full Gross-Pitaevskii equation \cite{sd96}
has a rounded-off tail 
which vanishes exponentially. In the large
$r$ region (near boundary) the density is very low and 
therefore the kinetic energy cannot be any longer neglected compared to the 
interaction energy term ($\equiv \rho(\bs{r})^2$) in the expression 
(\ref{gpen1}). The modification due to this non-negligible kinetic
energy will be discussed briefly in a later section (see \S\ref{ssurf}).

\setcounter{equation}{0}%
\renewcommand{\theequation}{\mbox{\arabic{section}.\arabic{equation}}}
\section{Vortex solution in the Thomas-Fermi regime}
We shall now study the vortex solution under the TF expression. 
We consider the case where the trap has an axis of symmetry,  
namely $\omega_x = 
\omega_y =\omega_{\perp}$ (\S\ref{2DIO}). In such a geometry
the TF radius in the $x-y$ plane is given by $R_x~=R_y~=R_{\perp}$. 
The condensate wavefunction in this case is of the form
(see section \S\ref{V})
\beq
\P = |\P (\bs{r})| e^{i S(\bs{r})}
\label{psi1expis}
\eeq
In the  
cylindrical polar co-ordinate  ($r, \theta, z$)
one again sets $S=\kappa \theta$, so that 
\begin{equation}
 \bs{v}_s =    {\hbar \over m r } \kappa\hat{\theta}  \; ,
\label{v}
\end{equation}
Using the harmonic oscillator length scales 
$a_{\perp}=1/\sqrt{b_{\omega_{\perp}}}$ and $a_z=1/\sqrt{b_z}$ 
and the harmonic 
oscillator energy scale $\hbar \omega_{\perp}$ first we write the 
GP equation (\ref{GPEQ}) in a dimensionless form 

\beq 
\left[-\frac{\hbar^2}{2m}\left[ \frac{\partial}{\partial r}(r 
\frac{\partial}{\partial r}) + \frac{\kappa^2}{r^2} \right]+\frac{1}{2}m
\omega_{\perp}^2 (r^2 + \lambda_R^2 z^2) + g|\P(\bs{r})|^2\right]\P(\bs{r})
=\mu \P(\bs{r}) \label{vortexnlse}\eeq

Due to the presence of the centrifugal term, the solution of this
equation  for $\kappa \ne 0$ has to vanish on the $z$-axis. From
the asymptotic behaviour of the Eq. (\ref{vortexnlse}) 
near the origin it can
be verified that the region over which the density changes very fast
to zero is given by the healing length $\xi_{0}$. 
Therefore in this region the kinetic energy cannot be neglected. 
Away from the origin the density profile is going to be same 
as the TF density without a vortex.
The TF approximation can now be applied by neglecting the radial part
of the kinetic energy (since it is sub-dominant relative to the centrifugal
term for small $r$) which gives the following TF density profile
\cite{SF98}  
\beq \rho_{\kappa} \approx
 \rho(0)\left( 1-\frac{\kappa^2\xi_0 ^2}{r^2}-%
\frac{r^2}{R_{\perp }^2}-\frac{z^2}{R_z^2}\right) \Theta \left( 1-%
\frac{\kappa^2\xi_0 ^2}{r^2}-\frac{r^2}{R_{\perp }^2}-\frac{z^2}{%
R_z^2}\right) \label{TFVD} \eeq
where we have used the relation (\ref{hlmu}) assuming that the 
correction to the $\mu_{TF}$ due to the presence of a vortex can be 
neglected \cite{Sinh97}. Thus the density profile of a vortex in TF
approximation significantly differs from the ground state density 
profile only over a length scale $\xi_0$.  

\subsection{Critical frequency of vortex nucleation in the TF approximation}
The calculation for the critical frequency under 
TF approximation was carried on by Lundh {\it et. al.}
\cite{Lund97}, Sinha \cite{Sinh97} and Feder \cite{F99}. 
The following discussion closely follows \cite{Lund97}.
The method consists in evaluating the extra energy associated 
with the presence of a vortex 
and then divide it by the total angular momentum, both 
determined within the TF approximation. The ratio then gives 
the critical frequency. 
Once there is a finite confinement along the $z$ direction
the system is not exactly like a infinite cylinder. 
But if the aspect ratio (\ref{AR}) $\lambda_{R} \ll 1$, 
a cylindrical profile of the trap is still a good approximation.
 
First one assumes that the TF radius satisfies the 
condition $R_{\perp} \ll R_{z}$. 
To obtain the energy of a vortex, Lundh {\it et. al.} \cite{Lund97} proposes that for such a large condensate
the domain of integration in the transverse 
plane can be divided into two parts.
In the region having a radius $r_{i}$ such that 
$\xi \ll r_{i} \ll R_{\perp}$ one can use 
the expression (\ref{GINZPITA}) after replacing $b$ for 
$r_{i}$ and for the region 
$ r_{i} < r < R_{\perp}$ one obtains the energy by 
integrating the kinetic energy term (interaction is neglected since the 
density is low in this region).
This gives the energy of a vortex ($\kappa=1$) per unit length as  
\bea E_{v}&=&\pi \rho(0)\frac{\hbar^2}{m}
\log\left( 1.464 \frac{r_{i}}{\xi_{0}}\right) 
+\frac{1}{2}\int_{r_{i}}^{R_{\perp}}2 \pi r dr  m \rho(r)v_{s}^2(r)\nonumber \\
~& \approx &~\pi \rho(0)\frac{\hbar^2}{m}\log\left(\frac{0.888R_{\perp}}
{\xi_{0}} \right) \label{ventf2d}\eea
In the second integral we use the ground state TF density 
(\ref{TFDP}) since in this domain it is $\approx \rho_1$ (\ref{TFVD}). 

The total angular momentum for a vortex state with $\kappa=1$  
in the TF approximation is given by 
\beq L_{z}=\hbar\left[\rho(0)\int_{0}^{R_{\perp}}
(1- \frac{r^2}{R_{\perp}^2})2\pi r dr \right]= 
\frac{1}{2}\hbar \rho(0)\pi R_{\perp}^2  \eeq

The critical frequency is therefore 
\beq \Omega_{c1}=2\frac{\hbar}{mR_{\perp}^2}
\log\left(\frac{0.888R_{\perp}}{\xi_{0}}\right) 
\label{crittf}\eeq 

Energy of the condensate with a vortex is always 
measured relatively to the energy of the bare 
condensate. The energies due to the harmonic confinement
($\frac{1}{2}m\omega_{\perp}^2 r^2 \rho$)
for these two cases differ appreciably from each other
for $r < \xi_0$. Hence its contribution is neglected in the 
above derivation.  
 
The next correction to the formula of the characteristic 
nucleation frequency comes by assuming the TF radius along the 
$z$ direction 
is finite. As mentioned earlier the shape of the cloud in this case 
is that of an 
ellipsoid and as a result its TF radius in the $x-y$ plane 
is $z$ dependent. The energy for a vortex in this case can be obtained by 
integrating the 
energy of a vortex state per unit length in two-dimension 
(\ref{ventf2d}) along the $z$-axis
from $-R_{z}$ to $R_{z}$ after replacing 
$R_{\perp}$ by $R_{\perp}(z)=R_{\perp}(1-\frac{z^2}{R_{z}^2})$. 
Such a calculation 
yields the following expressions: 
\beq E_{v}=\rho(0)\frac{\pi \hbar^2}{m}\int_{-R_{z}}^
{R_{z}}dz(1-\frac{z^2}{R_z^2})
\log\left(\frac{0.888R_{\perp}}{\xi_{0}}
(1-\frac{z^2}{R_{z}^2})\right)=\frac{4\pi \rho(0)}{3}
\frac{\hbar^2}{m}R_{z}\log\left(\frac{0.671R_{\perp}}{\xi_{0}}\right) 
\label{lundven} \eeq
\beq L_{z}=\hbar\left[\rho(0)\int_{-R_{z}}^{R_{z}}\left(1- 
\frac{r^2}{R_{\perp}^2} - \frac{z^2}{R_z^2}\right)dz 2\pi r dr \right]= 
\frac{8\pi}{15}\hbar \rho(0)\pi R_{\perp}^2R_{z}  \eeq
\beq \Omega_{c1}=\frac{E_{v}}{L_{z}}=
\frac{5}{2}\frac{\hbar}{mR_{\perp}^2}
\log\left(\frac{0.671 R_{\perp}}{\xi_{0}}\right) 
\label{lundvcrit}\eeq
Within this approximation 
it is important to note that for such an axisymmetric vortex 
the critical frequency 
does not depend on the TF radius along the axis of symmetry, 
though however the change 
in the shape of the cloud influences 
the absolute value of the critical frequency.  
This ends our discussion about the static properties of a TF 
(large) condensate.
We shall now switch over to the description of the dynamic extension
of this approximation.

\setcounter{equation}{0}%
\renewcommand{\theequation}{\mbox{\arabic{section}.\arabic{equation}}}

\section{Hydrodynamic theory of the condensate}\label{subhyd}
The study of a large (or equivalently strongly interacting) 
condensate can be extended to the time-dependent
situations by formulating a hydrodynamic theory  
of the condensate. 
The hydrodynamic equations \cite{SS96}
describes the time evolution of the density and the velocity field
of the condensate. Vortex nucleation in a rotating trapped condensate
is a dynamic process. Therefore the time-evolution of the density and
the associated velocity field plays a pivotal role in 
the nucleation mechanism. For a 
large and strongly interacting condensate (TF limit)
these hydrodynamic equations
are identical to the Euler equations of the classical hydrodynamics. 
A linear stability analysis of the hydrodynamic equations
of the condensate \cite{SS96}
gives a description of the collective excitations 
in terms of the density fluctuations and the associated velocity
fields. 

We shall see in the following discussion that one 
actually gets an 
analytical description of this collective excitations
using this classical hydrodynamic theory. Since the Bogoliubov
theory also describes the collective excitations one may ask 
how the collective excitations of the 
the hydrodynamic theory are related to the Bogoliubov
excitations. 
Applying the TF approximation directly to the 
Bogoliubov equations through a large $N_0$ expansion Sinha \cite{Sinh97}
(see also \cite{SF98}) showed that the
dispersion law \cite{SS96} for   
the collective excitations obtained under classical 
hydrodynamic approximation can be reproduced. 
The hydrodynamic theory of the condensate 
has been discussed in a number of references and has also been 
reviewed in \cite{Dalf99, fsr01, CastinR, book}. 
We discuss it in the following section briefly.

\subsection{Time evolution of the condensate density and velocity}
\label{subhy1}
To enable the construction of a hydrodynamic theory of the condensate 
the wavefunction 
$\Psi(\bs{r})$ can be split into a modulus and a phase 
\beq \Psi(\bs{r}) = \sqrt{\rho({\bs{r}})}e^{i\frac{S(\bs{r})}{\hbar}} \eeq
and the condensate energy (\ref{gpen1}) rewrites as 
\beq
E_{GP}[\rho,S]  = \int d \bs{r}\; 
\left[\frac{\hbar^2}{2m}\left(\bs{\nabla}\sqrt{\rho}\right)^2
+\rho\frac{\left(\bs{\nabla}S\right)^2}{2m} + V_{tr}\rho +\frac{g}{2}\rho^2
\right]
\eeq
Note that here we have explicitly divided the condensate phase by $\hbar$.
The corresponding Lagrangian is   
\beq
L_{GP} = -\left[\rho\partial_t S +\frac{\hbar^2}{2m} 
\left(\bs{\nabla}\sqrt{\rho}\right)^2
+\rho\frac{\left(\bs{\nabla}\,S\right)^2}{2m}+ V_{tr}\rho +
\frac{g}{2}\rho^2\right].
\eeq
Treating  $\rho(\bs{r},t)$ and $S(\bs{r},t)$ as 
generalized co-ordinates, one gets respectively the following    
Euler-Lagrange equations that define the full hydrodynamic
theory of the condensate
\bea
\partial_t S +\frac{1}{2m}\left(\bs{\nabla}S\right)^2+ V_{tr}+\rho g &=&
\frac{\hbar^2}{2m} \frac{\nabla^2\sqrt{\rho}}{\sqrt{\rho}}.
\label{hydro1} \\
\partial_t\rho + \bs{\nabla}\cdot\left[\rho\,\bs{v_s}\,\right]=0.
\label{conti}
\eea
with 
\beq \bs{v}_{s}=
\frac{\bs{j}_{\mbox{\scriptsize proba}}}{\rho} = 
\frac{\hbar}{2im}\left[\P^{\ast}\bs
{\nabla}\P
-\mbox{c.c.}\right] = \frac{\bs{\nabla}S}{m} 
\eeq
is  the local velocity field of the condensate.
The equation (\ref{conti}) is the usual continuity equation. 
The time evolution of the velocity field is obtained by
taking the gradient of both sides of the equation (\ref{hydro1}).
\bea
m\partial_t\bs{v_s}&=&-\bs{\nabla}\left[\frac{1}{2}mv_s^2 +
V_{tr}(\bs{r}\,) + 
g\rho(\bs{r}\,)
-\frac{\hbar^2}{2m}\frac{\nabla^2\sqrt{\rho}}{\sqrt{\rho}}\right] \nonumber \\
&=&-\bs{\nabla}(\frac{1}{2}mv_{s}^2+\mu) \label{evolvo} \eea
where the chemical potential is defined as  
\beq \mu=V_{tr} + g\rho -
\frac{\hbar^2}{2m}\frac{\nabla^2\sqrt{\rho}}{\sqrt{\rho}} \label{cpquant} \eeq
Using the expression (\ref{chemtf}) the above equation can be 
rewritten as 
\beq \mu = \mu_{TF} -
\frac{\hbar^2}{2m}\frac{\nabla^2\sqrt{\rho}}{\sqrt{\rho}} \label{qp1}\eeq  
For a uniform BEC ($V_{tr}=0$)
at $T=0$, the chemical potential
$\mu = g\rho_{0}$ and the pressure $p=-\frac{\partial E}{\partial V}|_{T}=
\frac{g}{2}\rho_{0}^2$. Using the same expression for pressure after 
replacing $\rho_0$ by $\rho$, (\ref{evolvo}) can be rewritten as 
\beq m\partial_t\bs{v_{s}}=-\bs{\nabla}\left[\frac{1}{2}mv_s^2 +
V_{tr}(\bs{r}\,) + 
\frac{\bs{\nabla}p}{\rho}
-\frac{\hbar^2}{2m}\frac{\nabla^2\sqrt{\rho}}{\sqrt{\rho}} 
\right] \label{hydr21}\eeq
which is very similar to the Euler equation for classical
hydrodynamics 
\beq m\partial_t\bs{v_{s}}=-\bs{\nabla}\left[  m(\bs{v}_s\cdot\bs{\nabla})
\bs{v_s}+
V_{tr}(\bs{r}\,) + 
\frac{\bs{\nabla}p}{\rho}\right] \eeq
where we use 
$\bs{\nabla}(\frac{1}{2}mv_{s}^2)= m(\bs{v}_s\cdot\bs{\nabla})\bs{v_s}$. 
The only exception is the so called quantum pressure term, given by 
\beq \mu - \mu_{TF}=-\frac{\hbar^2}{2m}\frac{\nabla^2\sqrt{\rho}}{\sqrt{\rho}},
\label{qp}
\eeq
the only term in the equations 
where $\hbar$ appears. The source of this term is the kinetic energy
of the condensate. 

\subsection{Classical hydrodynamic approximation}\label{subhy2} 
If the scale of variation of $\rho$ is of the order of $R$, the 
pressure term in the equation (\ref{hydr21}) is of the order 
$\frac{g\rho}{mR}$ whereas the quantum pressure term (\ref{qp}) is 
of the order $\frac{\hbar^2}{m^2R^3}$. Therefore the 
quantum pressure term 
can be neglected as long as $
\frac{(\frac{\hbar^2}{m^2 R^3})}{(\frac{g\rho}{mR})} \ll 1$. This implies
$R \gg \frac{\hbar}{\sqrt{m g\rho}}$ which finally gives the condition
\beq R \gg \xi \eeq
This condition is satisfied for 
a large TF condensate (\ref{tflength}). 
In this limit neglecting the quantum pressure term 
in (\ref{hydr21})
we get the classical hydrodynamic equation for the condensate  
 \beq m\left(\partial_t+\vec{v}\cdot\bs{\nabla}\right)\vec{v} =
-\bs{\nabla}[V_{tr}+g\rho].
\label{classhydro}\eeq
and the chemical potential $\mu$  becomes  $\mu_{TF}$.  

Finally we write the 
classical hydrodynamic equations in a trap 
rotating with an angular velocity $\bs{\Omega}$
\beq
\partial_t\rho + \bs{\nabla}\cdot\left[\rho\,(\bs{v_s} - 
\bs{\Omega}\times \bs{r})
\,\right] = 0 \label{rothydr1}\eeq
\beq 
m\partial_t\bs{v_{s}} +\bs{\nabla}\left[\frac{1}{2}mv_s^2 +
V_{trap}(\bs{r}\,) + 
\frac{\bs{\nabla}p}{\rho} - \bs{v}_s.(\bs{\Omega} \times \bs{r})
\right] = 0 \label{rothydr2}\eeq

\subsection{Collective excitations from the classical 
hydrodynamic theory}\label{colhydro}
Using the hydrodynamic equations (\ref{conti}) and (\ref{classhydro}) 
Stringari \cite{SS96} studied the  
collective excitations of a trapped condensate. For a review see
also \cite{book, book1, Dalf99, CastinR}.
Let us denote the steady state (time-independent) density and velocity 
satisfying  (\ref{conti}) and (\ref{classhydro}) by $\rho_s$ and $0$.
We follow the method described in  the section 
(\S\ref{GPLR}) in order to linearize these equations about 
the steady state solutions. This yields
\bea
\frac{\partial \delta\rho}{\partial t}+\bs{\nabla}\cdot
\left[\rho_s\delta\bs{v}_{s}\,\right] &=& 0 \label{linhydro1} \\
m\frac{\partial \delta\bs{v}_s}{\partial t} +
\bs{\nabla} [\delta\rho g] &=& 0 
\label{linhydro2}
\eea
which gives 
\beq
\partial_t^2\delta\rho - \bs{\nabla}\cdot\left[c_s^2(\bs{r}\,)\bs{
\nabla}\,\delta\rho\right]=0 
\label{1linhydro}\eeq
while defining  $mc_s^2(\vec{r})= \rho_s(\bs{r})g$. 
Therefore in a non-uniform trapped condensate 
the collective excitations 
which take the form of density fluctuations 
propagates as sound waves with a position
dependent sound velocity $c_s(\vec{r}\,)$ unlike the collective excitations 
in a uniform condensate where such a sound velocity is constant
(\ref{soundvel}). 
The collective mode frequencies can be obtained by writing
\beq \delta \rho(\bs{r},t)=\delta \rho(\bs{r})e^{i\omega_{cl}t} \eeq
Since the steady state density  is same as the TF density, {\it i.e.} 
$\rho_s= \frac{\mu_{TF} - V_{tr}(\bs{r})}{g}$, 
the eigenvalue equations are 
\beq
-\omega_{cl}^2\delta\rho=\frac{1}{m}\left[\bs{\nabla}V_{tr}(\bs{r}).
\bs{\nabla}
\delta \rho - (\mu_{TF} - V_{tr}(\bs{r}))\nabla^2 \delta \rho
\right] 
\label{1linhydro} \eeq 
The solutions of these modes have been obtained by Stringari \cite{SS96}. 
Here we mention some 
results which will be later 
used to explain the process of vortex nucleation 
in the current experiments.

\subsection{Hydrodynamic Modes for trapped condensates}\label{HMTC}
\begin{itemize}
\item Spherical trap
\end{itemize}
For the spherical trap described in (\S\ref{subsb})  
Eq. (\ref{1linhydro}) becomes
\beq -\omega_{cl}^2\delta\rho=\left[\omega^2r
\frac{\partial}{\partial r}
\delta \rho - \frac{\omega^2}{2}(R^2 - r^2)\nabla^2 \delta \rho
\right] \label{lhdsph} \eeq 
where $R$ is the TF radius (\ref{tfbc}). 
The  eigenvalues are given by \cite{SS96}
\beq
\omega_{cl}^2 = (l + 3n_{r} + 2n_{r}l + 2n_{r}^2)\omega^{2} 
\label{hydrodisp} \eeq
The normal modes of the density fluctuations are given 
in terms of the hypergeometric functions and the spherical harmonics, namely 
\beq \delta \rho(\bs{r}, t)= C r^{l}F(-n_{r},l + n_{r} + 
\frac{3}{2}, l + \frac{3}{2} ; \frac{r^2}{R^2})Y_{l, l_z}(\theta, \phi)
e^{ -i 
\omega_{cl} t} \label{hydrodens} \eeq
Details of this calculation is also available in chapter ${\rm 7}$ of 
ref. \cite{book}.
Here $n_{r}$ gives the number of radial nodes and $l$ gives the polarity 
of the density oscillation.
This result can be compared with the dispersion relation for noninteracting
bosons in a spherically symmetric trap (\ref{sphe}) which is 
(after subtracting the ground state 
energy)
\begin{equation}
\omega_{cl} (n_r,l) = \omega(2n_r + l)
\label{shosph}
\end{equation}

The modes with no radial nodes  ($n_r=0$) are called the 
{\bf surface excitations} for which  (\ref{hydrodisp}) predicts
the dispersion law  
\beq \omega_{cl} = \sqrt{l}\ \omega \label{surfe}\eeq.
The frequency of these modes is  smaller than the harmonic
oscillator result $l\omega$ (\ref{sphe})
except for the case of $l=1$
mode (dipole mode, for details see \cite{FR98}). 
At very low energy only these modes are excited. Therefore
when the condensate is rotated, initially all the angular momentum
are carried by these modes. Their time evolution under rotation actually
determines how angular momentum is transferred from the surface 
to the bulk of the systems and leads to the nucleation of a vortex. Specially 
the surface 
mode with $l=2$ and energy $\sqrt{2}\omega$, 
known as the quadrupole mode, plays a very prominent role 
in the vortex nucleation mechanism. 
Experimental 
traps are generally non-spherical but has an axis of symmetry. So 
we identify such surface modes in this geometry in the next section.

\begin{itemize}
\item{Axisymmetric trap}
\end{itemize}
In
an axisymmetric confinement (\S\ref{2DIO})
Eq. (\ref{1linhydro}) writes as 
\beq \omega_{cl}^2\delta \rho = \omega_{\perp}^2
\left( r\frac{\partial}{\partial r}
+ \lambda_{R}^2z\frac{\partial}{\partial z} \right)\delta \rho 
-\frac{\omega_{\perp}^2}{2}
(R_{\perp}^2 - r^2 - \lambda_{R}^2z^2)\nabla^2\delta\rho \eeq
where $\lambda_{R}$ is the aspect ratio and 
$R_{x}=R_{y}=(1/\lambda_R) R_{z}=R_{\perp}$. 
Unlike the case of a spherical trap  
here all density fluctuation
modes and their dispersion cannot be written 
in an analytic form. However, using the symmetry 
about the $z$-axis some specific low lying 
modes can be studied. Because of this symmetry
the normal modes of density fluctuations 
are eigenmodes of the operator $L_{z}=\frac{\hbar}{i}\frac{\partial}
{\partial \theta}$. We use $\theta$ as polar angle in the spherical polar
co-ordinate system and as azimuthal angle 
in the cylindrical polar co-ordinate system.
Also for notational convenience here we
denote $l_z=\pm p$ where $p \in \bs{N}$.
One class of such solutions is of the form 
\beq \delta \rho \propto z^{p}=(x+iy)^{p}= 
r^{p}Y_{p,\pm p}(\theta, \phi) \label{qdp11} \eeq
which  have the dispersion  
\beq \omega_{cl}^2 = p \omega^2 \label{qdp12} \eeq
Therefore they can be identified as the surface modes. 
Similarly it can be verified that there are solutions of the 
form 
\beq \delta \rho \propto z(x+iy)^{p-1} =r^{p}Y_{p,\pm (p-1)}
\label{qdp21} \eeq
with the frequencies given by 
\beq \omega_{cl}^2 = (p -1)\omega_{\perp}^2 + \omega_z^2 \label{qdp22} \eeq
These two modes are degenerate for a spherical trap 
and both of them surface modes.

To identify the quadrupole modes we note that
for $\lambda_{R} \neq 1$, there is no spherical symmetry
and as a result $l$ is not a good quantum number. Since $l$ determines 
the polarity, determining polarity of a general mode in this trap is 
difficult. But as shown in (\ref{qdp11})
and (\ref{qdp21}) there are eigenmodes in such a system which can be 
mapped on to the eigenmodes of the condensate in a spherically
symmetric  confinement and their polarity can be subsequently identified.
In this way one gets following two expressions for the quadrupole density
modulation from (\ref{qdp11}) and (\ref{qdp21}) \cite{SZ98} after setting 
$p=2$, namely 
\bea \delta \rho \propto r^2e^{\pm 2 i \theta} &=& (x +iy)^2,~~ l_z=\pm 2 
\label{qp1} \\
\delta \rho \propto rze^{\pm i \theta}&=&z(x+iy),~~l_z=\pm 1 \label{qp2} \eea
with their frequencies given by (\ref{qdp12}) and (\ref{qdp22}).
Let us consider a linear combination of the two modes described in 
(\ref{qp1}) of the form 
\beq \delta \rho \propto (x^2 - y^2) \label{quaddens} \eeq 
Since this mode is a 
superposition of two eigenmodes with 
equal and opposite $l_z$, its angular momentum about
the axis of symmetry is zero. If such a mode becomes the 
steady state solution
of the system, 
using (\ref{conti}) it can be shown 
that the associated quadrupolar velocity field $\bs{v}_s$ is 
given as $\propto \bs{\nabla}(xy)$ 
Since the angular momentum about the $z$-axis is $0$, such a field does
not contain any vorticity. However if the trap is rotated 
new situation arises \cite{RZS01}.
We discuss this in a later section.

\subsection{Quadrupole modes in the presence of a vortex}\label{quadv}
In the earlier section it has been pointed out that 
if the confinement is isotropic in the $x$-$y$ plane, 
two quadrupole modes having angular momentum $\pm l_z$  
are energetically degenerate. Though this is true for 
a condensate without vortices, in the presence of a vortex
this degeneracy is lifted \cite{Sinh97, SF98}. 
This can be explained by noting that the average velocity
flow associated with the collective oscillation can be either 
parallel or  opposite to the vortex flow, depending on the 
sign of the angular momentum carried by  the excitation. This 
provide a chance to spectroscopically detect the presence of a vortex
in the condensate. To that purpose Zambelli and Stringari \cite{SZ98},
using  a sum rule based approach, 
derived analytical expression which 
relates the splitting between the quadrupole modes with the average 
angular momentum per particle. For the details of the sum-rule approach 
see \cite{SL89} and how this is applied to study the 
response of the quadrupole mode
is given in detail in \cite{SZ98}. We shall just mention the expression
obtained by Zambelli and Stringari  which relates 
how the frequency difference between  
two quadrupole modes
with equal and opposite angular momentum $l_{z}$ is associated with 
the average angular momentum per particle in the superfluid. 
Within the sum-rule approach 
the
splitting between the two frequencies can be written as

\beq
\omega_{+2}\!\!-\omega_{-2}\!=\frac{2}{m}
\frac{\langle L_z\rangle}
{\langle r^2_{\perp} \rangle}=\frac{7\omega_{\perp}\kappa }{
\lambda_R^{2/5}}\bigg(\!15\frac{N_0a}{a_{\perp}}\bigg)^{-2/5}
\label{hdshift}
\eeq
where $\omega_{+2}$ and $\omega_{-2}$ correspond to two quadrupole modes 
with $l_z\!=\!\pm 2$ and $a_{\perp}=\sqrt{\frac{\hbar}{m \omega_{\perp}}}$. 
This result has been subsequently used by the ENS \cite{CMD00}
and JILA \cite{JILA2} group
to measure the
angular momentum of the condensate in the presence of a vortex.

\setcounter{equation}{0}%
\renewcommand{\theequation}{\mbox{\arabic{section}.\arabic{equation}}}%
\section{Experiments on vortices in BEC}
In this section we describe the experiments done to detect
the vortices and to study their properties in the atomic BEC. We also compare 
the results of these experiments with the theoretical predictions 
which have been described in the earlier sections.

The first experimental observation of vortices in a trapped  
condensate took place in JILA \cite{JILA1}. The 
experimental scheme is guided by a theoretical model 
 proposed by Williams and Holland \cite{WH99}
which exploits
the possibility of trapping otherwise identical (bosonic) atoms in two
different internal (hyperfine) states (chapter $3$ of \cite{book}). 
The basic idea \cite{WH99} is as follows. 
Let us assume that identical atoms in two hyperfine states are trapped
in two identical axially symmetric harmonic oscillator potentials, whose
centers are spatially separated by a given distance. These two traps are
rotated about the common axis of symmetry which passes through the center
of the line joining the centers of these traps, with a frequency $\Omega$.
Simultaneously an electromagnetic field is applied that couples the two
internal atomic hyperfine states. The coupling is parametrized by 
two parameters, namely the detuning $\delta_{dt}$ and the Rabi frequency
${\cal R}$. The detuning $\delta_{dt}$ gives the mismatch of the frequency 
of the coupling
electromagnetic field to the frequency (energy) difference between the two
hyperfine states. And ${\cal R}$ gives the rate at which the population
would oscillate between these two states if  $\delta_{dt}=0$. 

The condensate
wavefunction of this two-component BEC can be written as a two-component 
spinor. The  symmetry of such wavefunction is different from 
that of a single-component BEC wavefunction \cite{fsr01}. 
For explaining the basic idea behind
the vortex nucleation in this scheme, it is enough
to consider the following part of the free energy functional in the 
co-rotating frame
\beq F_{part}(rot)=\int d\bs{r}\sum_{i=1,2}\P_i^{\ast}
(H_{0} - \Omega L_{z})\P_i + \frac{\hbar}{2}\delta_{td}
(\P_1^{\ast}\P_1 - \P_2^{\ast}\P_2) \eeq
where $1$ and $2$ denotes two hyperfine states.
We know that(\S\ref{gpcritf})
the energy of a vortex with one unit of angular momentum 
($\kappa=1$) is
shifted by $\hbar \Omega$ in the co-rotating frame relative to its
energy in the laboratory frame. When this energy shift is compensated
by the energy mismatch $\hbar\delta_{dt}$ due to detuning, one gets 
one condensate
at rest in the center of the atomic cloud and the other in a unit vortex
state around it. This has been realized in JILA \cite{JILA1}.
Subsequently the central region can be removed and one
gets a vortex state in a single-component BEC \cite{JILA3}.
Apart from these experiments in JILA the other experiments to detect
vortices is analogous to the Rotating Bucket experiment for the 
Superfluid $^4$He \cite{Don91}. In this experiment, at very low temperature 
the sperfluid helium 
will
come to the rest in the frame of a rotating bucket when the rotational
frequency reaches a characteristic value. At this frequency a
quantized vortex is nucleated in the superfluid.
We discuss similar experiments done for the trapped 
condensate in the following section.

\subsection{Rotating Bucket analogue for trapped atoms}\label{ENS}
For the trapped atom, the role of the bucket can be played by
the trapping potential. Therefore a similar experimental
set up may be constructed if one can rotate the trap in which 
the condensate is formed.
Such an experimental set-up 
to detect vortices was first realized by  
the ENS group \cite{MCWD99, MCWD00}
where the the trap is put into rotation 
by stirring it with a laser beam. This leads to the 
observation of the vortex nucleation in the rotating trap.
Subsequently employing the same technique MIT group \cite{MIT1, MIT2} 
also observed the nucleation of a 
large number of vortices. Vortices nucleated in this way arrange 
them in the form of a vortex lattice \cite{Abrikosov}.  
In the experiment done by the Oxford
group \cite{OXFORD} the trap is  
rotated by using magnetic fields instead
of laser stirrer. Here we describe
how the trap is rotated by using a stirring laser beam 
following \cite{CMBD01}.

\subsection{Experimental realization of a rotating trap}\label{rottrap}
The magnetic trap in which the the BEC is confined 
is an axisymmetric  harmonic potential:
\beq
V_{tr}({\bf r})=\frac{1}{2}m\omega_{\perp}^2(x^2+y^2) + 
\frac{1}{2}m\omega_z^2 z^2,
\end{equation}
If the aspect ratio (\ref{AR}) $\lambda_{R} \ll 1$ the shape of 
the condensate 
is like a cigar with its height
(TF radius in $z$-direction) is much larger than its
diameter (TF radius in the transverse plane) \cite{CMBD01, MIT1}.
In the other limit ($\lambda_{R} \gg 1$) the condensate is like
a pancake \cite{JILA2}. 
To create a rotating trap 
the condensate is stirred with a focused 
laser beam propagating  along the symmetry axis ($z$-axis)
of the trap and toggling 
back and forth very rapidly between two symmetric positions 
about the center of the trap. 
The electromagnetic field of the laser beam
creates an average dipole potential in the $x-y$ plane 
which gives the following additional effective confinement 
\beq
\delta V_{tr}({\bf r})=\frac{1}{2}m\omega_{\perp}^2
(\epsilon_X X^2+\epsilon_Y Y^2)
\label{potential}\eeq
where $X$ an $Y$ are the co-ordinates in the co-rotating frame. 
This potential is rotated through an 
acousto-optic 
deflector which rotates the co-ordinates $X$ and $Y$ at an angular frequency
$\Omega$ producing a rotating harmonic trap. Therefore
\beq
X=x\cos (\Omega t)+y\sin(\Omega t)\qquad
Y=-x\sin(\Omega t)+y\cos(\Omega t) \label{XY}
\eeq
Defining $\omega^2_{X, Y}=\omega_{\perp}^2(1 + \epsilon_{X,Y})$, 
the total confinement potential in the lab frame is given by 
\begin{eqnarray}
\left(V_{tr}+\delta V_{tr} \right)(\bf r) &=& \frac{1}{2}m
(\omega_X^2 X^2 +\omega_Y^2 Y^2)+\frac{1}{2}m\omega_z^2 z^2
\label{corot}
\\
&=& \frac{1}{2}m\bar \omega_{\perp}^2(x^2 +y^2)+\frac{1}{2}\lambda_R^2
m\omega_{\perp}^2
z^2 \nonumber \\
&& +\frac{1}{2}m \,\epsilon \,\bar\omega_\perp^2
\left((x^2-y^2)\cos(2\Omega t)+2xy\sin(2\Omega t)  \right)
\label{lab}
\end{eqnarray}
with 
\beq
\overline{\omega} = \sqrt{(\omega_{X}^2+\omega_{Y}^2)/2}\qquad
\qquad
\epsilon = (\omega_{X}^{2} - \omega_{Y}^{2})/(\omega_{X}^{2} +
\omega_{Y}^{2})\ .
\label{anisotropy}\eeq
This potential is stationary in the rotating frame 
and oscillates periodically at frequency
$2\Omega$ in the laboratory frame (\ref{lab}).
The stirring frequency $\Omega$ is chosen in the
interval $(0,\omega_{\perp})$ so that the condensate is stable. 
At the upper value of this interval,
the centrifugal force equals the transverse restoring force of
the trap.
Similar technique to rotate the trap
is employed in the MIT experiments with a varying range 
of intensity and frequency of the laser stirrer. Particularly 
the anisotropy $\epsilon$ in their trap is higher in comparison to
the other experiments.  
The Oxford group \cite{OXFORD} 
puts the atom in a combination of the
spherical quadrupole magnetic field and a rapidly rotating bias field
(magnetic). Together they form a time-averaged orbiting potential
which is anisotropic in the transverse plane and also rotates the trap.

\subsection{How to detect a vortex?}
\begin{itemize}
\item{\bf From the optical image of the condensate density}
\end{itemize}
In the rotating trap set-up
the vortices will be nucleated when the stirring frequency 
exceeds a characteristic value $\Omega_{c1}$. Usually 
the stirring frequency is equal to the frequency at which the
trap is rotated.
The TF density of the  condensate
with a vortex (\ref{TFVD}) vanishes at the 
center of the trap (vortex core).
Therefore one may detect a vortex through the imaging 
of the density profile of the condensate. 
To this purpose the 
stirring potential is switched off adiabatically (in a time
long compared to $\omega_{\perp}^{-1}$) and the
condensate density profile along the stirring axis
is imaged to detect the presence of a vortex.
The 
radius of the vortex core is of the order of the  healing
length (\ref{hlmu}) $\xi_0=(8\pi \rho(0) a)^{-1/2}$ which is too small
(like $\xi_0\sim 0.2\;\mu$m in the experiments of the ENS group) to
be observed optically in the experiments. 
To remove this difficulty the time of flight technique is used
in which the trap is switched-off and the condensate is allowed  
to expand for some time \cite{TFLT}. In this process the vortex core also 
gets expanded. The image of the enlarged condensate is then 
taken to detect the presence of a vortex. 

\begin{itemize}
\item {\bf Measurement of the angular momentum}
\end{itemize}
In the presence of a single vortex with winding number 
$\kappa=1$ each particle in the superfluid carries 
on the average extra one unit ($\hbar$)
of angular momentum. Therefore 
the presence of a vortex can be verified by measuring this 
extra angular momentum.
Both ENS \cite{CMD00} and JILA \cite{JILA2} did this  
measurement to confirm the presence of a vortex in a rotating condensate. 
For this they used the result (\ref{hdshift})
obtained by Stringari and 
Zambelli \cite{SZ98} that in a cylindrically symmetric geometry in 
the presence of a vortex , the difference between  
frequencies of the 
two transverse quadrupole modes $\omega_{+2}$ and $\omega_{-2}$,
 corresponding respectively to excitations with angular
momentum $l_{z}=2$ and $l_{z}=-2$ 
is proportional to the average angular momentum
per particle (\ref{hdshift}). 
In a static condensate these two modes are degenerate.
In the experiment \cite{CMD00}
the condensate is subjected to the 
dipole potential created by 
stirring laser beam but with a fixed basis ($\Omega=0$ in
Eq. \ref{XY}) such that 
$x= X$ and $y=Y$. Therefore one just have an elliptically 
deformed trap. This 
non-rotating dipole potential is applied for a period which is smaller
than the quadrupolar oscillation period. The condensate is then allowed
to oscillate freely in the pure magnetic trap.   
Using time-of-flight technique its image is then taken. This procedure was 
repeated in the presence as well as in the absence of a vortex. 
From the precessional frequency of the major and the minor axis 
of the ellipse the difference 
$\omega_{+2} - \omega_{-2}$ is determined. Using (\ref{hdshift}), 
the angular momentum per atom is measured from it.

\subsection{Vortex nucleation in a deformed trap}\label{rotdeform}
In a static trap the confinement potential in the
$x-y$ plane is isotropic. However once the rotation is switched on 
this is no more true since the stirring beam introduces an anisotropy 
in the confinement potential (\S\ref{rottrap}, Eq. \ref{potential}).
The anisotropy (\ref{anisotropy}) can be set either suddenly 
\cite{MCWD99, MIT1} or 
adiabatically \cite{MCBD01, OXFORD} resulting in 
an elliptic 
deformation of the condensate. This has important consequences for the 
process of vortex nucleation. 
ENS group \cite{MCBD01} made a detailed study of the stationary state 
of the condensate when this deformation is fixed at a given 
value and the rotational frequency is ramped up adiabatically. 
Their findings agree well with the theoretical prediction by 
Recati {\it et. al.}\cite{RZS01}. This stationary state 
becomes dynamically unstable beyond a characteristic rotational frequency. 
This leads to the nucleation of a vortex\cite{CS01}.
On the otherhand when the deformation is switched of suddenly  
the mechanism of vortex nucleation \cite{CS01, KPSZ01} 
is different.
We discuss this in a later section (\S\ref{detrap}).

\subsection{Rotating the normal cloud}
In another experiment by the JILA group \cite{JILA2}, 
instead of rotating the confining potential, 
the normal component of the atomic cloud which 
surrounds the superluid atomic cloud 
is rotated while the trap stays at rest. 
To observe vortex nucleation in this set-up the non-condensate atoms
which is in the normal state is
rotated first and then they are cooled below the condensation 
temperature. As a result the superfluid 
fraction grows from zero to  a finite value. However 
the boundary of the cloud is always formed of the normal-component and
rotates like a rigid body. The superfluid part 
in the interior rotates only
by creating a vortex.  
With increasing rotation
the surface of the 
cloud gets inflated in the plane of rotation because of the centrifugal
force. The rotational frequency can be measured from 
the change
in the size of the cloud in the plane of the rotation 
using the following expression
\beq \frac{\Omega}{\omega_{\perp}}=\sqrt{1 - 
(\frac{\lambda_{R}}{\lambda_{R}^0})^2}\eeq
where $\lambda_{R}^{0}$ is the aspect ratio (\ref{AR}) of the static trap.
The number of vortices in the 
superfluid is then determined from $\Omega$.
 
\subsection{How good  experiments agree with the theory?}
The comparison is done by comparing  
the theoretically determined (\ref{lundvcrit}) values 
of the characteristic frequency of 
the vortex nucleation \cite{Lund97, Sinh97, F99} to the 
corresponding experimental results. 
The critical frequency of vortex nucleation 
is dependent on the several parameters such as the number 
of particles in the condensate, stirring mechanism, anisotropy in
the trap potential, stirring time etc. 
Consequently the characteristic frequency in the vortex nucleation 
as a function of the transverse confinement frequency ($\omega_{\perp}$)
can vary in the experiments.  
This ratio is   $\approx 0.67$ in \cite{MCWD99} 
which is higher than the value 
calculated within TF approximation, ($\approx .41$) \cite{Lund97}. 
Moreover ENS experiments also do not show any strong dependence
on the number of condensate particles 
which is also in  contradiction to the 
TF prediction. For the parameters used in
the MIT experiment \cite{MIT1} 
the theoretical (\ref{lundvcrit})  rotational frequency 
of the first vortex nucleation 
$(\approx 0.08 \omega_{\perp})$ is also much smaller than the  
experimentally observed value $(\approx 0.3 \omega_{\perp})$.
In the JILA experiment \cite{JILA2}, the characteristic frequency
of vortex nucleation  is given by $0.32\omega_{\perp} 
< \Omega_{c1} < 0.38\omega_{\perp}$ which is also higher
than the TF estimation 
($0.2\omega_{\perp} < \Omega_{c1} < 0.25\omega_{\perp}$).
This is true also for the experiment done in Oxford
\cite{OXFORD}. Moreover according to the observations in ENS, MIT and Oxford 
the vortex nucleation in the BEC is always
preceded by a strong deformation of the surface of the atomic cloud. 
Particularly 
in the experiments by the ENS \cite{CMBD01} 
and the Oxford \cite{OXFORD} group this deformation 
has been identified as a 
surface quadrupole mode. Experiments 
\cite{MCBD01, OXFORD} also point out
mechanism of nucleation of a vortex is dependent on 
whether the trap is put into rotation adiabatically or suddenly. These
experiments motivate a more local approach to study the process of
vortex nucleation in a trapped BEC. Corresponding developments are 
discussed in subsequent sections.

\setcounter{equation}{0}%
\renewcommand{\theequation}{\mbox{\arabic{section}.\arabic{equation}}}
\section{The Role of surface barrier in vortex nucleation}
We have seen in the last section that the thermodynamic 
frequency of the vortex nucleation 
determined within the TF approximation (\ref{lundvcrit}) is usually
much smaller than the experimentally determined 
characteristic frequency of first vortex nucleation. This fact and the
other experimentally
observed features points out that the existence of 
the vortex state as a global minimum of the free energy of the condensate
in the rotating frame beyond 
a certain characteristic rotation 
frequency is a necessary but not the sufficient 
criterion for the nucleation of a vortex. The presence of a  
vortex in the bulk of a condensate in a static 
axisymmetric trap is associated with an extra energy.
Within the TF approximation it can be shown  
that this energy is highest when the vortex is centered at the center of the 
trap and decreases monotonically as a function of distance of the vortex
center from the center of the trap \cite{fsr01, KPSZ01, GG01}. 
Finally within TF approximation
(without any boundary correction) this energy vanishes at the boundary 
of the system. The vortices are therefore nucleated from the surfaces.
The dynamic 
evolution of the surface modes in a rotating trap and the change
in the energy landscape due to the presence of a vortex at a given
distance from the center of the trap determines the mechanism  
of vortex nucleation in a trapped BEC. 
This will be discussed in the following sections.
The discussion will start with 
the description of the Landau criterion of 
superfluidity \cite{LandauC}, which when 
applied to the surface modes in a trapped
condensate, determines at what rotational frequency such a mode will become 
energetically unstable towards the vortex nucleation in the 
center of a condensate.

\subsection{Landau criterion of superfluidity}
Landau criterion determines the condition for the superfluidity.
Let us  consider   
a particle of mass $m'$
 sent into the Bose-gas
with an initial velocity $\bs{v}_{p}$. The particle can transfer 
energy and momentum to the Bose-gas only by creating an excitation of 
momentum $\hbar \bs{k}$ which  causes viscous damping. If such
an excitation is produced, with momentum $\hbar \bs{k}$,  the 
momentum of the particle is reduced by the same amount $\hbar\bs{k}$. 
The conservation of energy therefore requires
\beq
E_{\bs{k}} = \frac{1}{2}m' \bs{v}_{p}\,^2-
\frac{1}{2m'} \left[m'\bs{v}_p-\hbar\bs{k}\,\right]^2= 
\hbar\bs{k}\cdot\bs{v}_p-
\frac{\hbar^2k^2}{2m'}.
\eeq
To produce a 
finite energy excitation, $|\bs{v}_p|$ has to satisfy the condition
\beq
|\bs{v}_{p}| \geq \left|\frac{\bs{k}\cdot\bs{v}_p}{k}\right| 
\geq \frac{E_{\bs{k}}}{k}
\geq c_s.
\eeq
The last expression follows from 
Bogoliubov dispersion law (\ref{bspectrum}).
This implies that 
a particle with an incoming velocity smaller than the sound velocity 
$c_{s}$ can move through the condensate without causing any damping. 
Therefore there is 
no channel at very low temperature through which the 
interacting Bose gas can dissipate. This leads to its superfluidity.
However, for an ideal Bose gas this condition gives 
\beq |\bs{v}_{p}| \geq \frac{E_{\bs{k}}}{k} \geq \frac{\hbar^2k}{2m}
\eeq
which makes it possible to excite
an ideal Bose gas with an infinitesimal $\bs{v}_p$.
Therefore it does not show superfluidity like a 
repulsively interacting Bose gas. Both systems however shows 
Bose-Einstein condensation.

We have already mentioned that
for a trapped condensate the 
low energy excitations are the surface Bogoliubov
modes (\ref{surfe}) which has finite angular momentum. Therefore 
the channel through which a trapped condensate can dissipate
is these surface excitations.
Dalfovo and Stringari \cite{DS00} showed how the generalization of the
Landau criterion can determine at what rotational frequency such 
a surface modes can be excited to nucleate a vortex in the condensate.
Let us denote the frequency of such a surface mode as $\omega_s(l)$. 
Then the  rotational frequency at which the $l$-th surface mode can be
excited is given by 
\beq \Omega(l) = \frac{\omega_{s}(l)}{l} \label{lcsurf}\eeq
This excited surface mode overcomes a potential barrier in the surface
and get in the bulk in the form of vortex.  
The minimum of the rotational frequency determined in this way
(\ref{lcsurf})  
therefore gives the characteristic frequency of first vortex
nucleation, namely $\Omega_{c1}$. 

Within classical hydrodynamical approximation (\ref{surfe}), 
the quantity $\frac{\omega_{cl}(l)}{l}$ monotonically decreases
with increasing $l$ and does not give a realistic value \cite{DS00}
for $\Omega_{c1}$. 
This is because the classical hydrodynamic approximation (\S\ref{subhy2})
is no more valid in the surface region where the kinetic energy
is appreciable and the interaction energy is relatively weak.
This problem was tackled in various ways. Dalfovo and Stringari used
\cite{DS00} a sum-rule based approach 
which fixes the upper limit
of the frequency $\omega_s(l)$. Khawaja {\it et. al.}\cite{AKPS99}
modified the classical 
hydrodynamic approximation 
by including the surface correction to the TF 
density \cite{DLS96} and obtain the dispersion law for various 
surface width. Their treatment was subsequently
improved by  Anglin \cite{JA01} through the use of a combination 
of analytical and 
numerical techniques and the determined critical frequency from the 
Landau criterion agrees well with the MIT experiment \cite
{MIT1}. Simula {\it et. al.}\cite{SMS} obtained the surface dispersion
relation by numerically solving Bogoliubov equations in a rotating frame.
Here we provide a brief discussion on the role of surface excitations
in the vortex nucleation process.

\subsection{The order-parameter at the surface of the cloud}\label{ssurf}
To find out the dispersion relation of the surface modes one needs to 
to take into account the modification of 
the TF approximation (\S\ref{subtfa}) near 
the surface of a condensate. This was carried out by
Dalfovo {\it et. al.}
\cite{DLS96}.
They started by identifying 
the characteristic length scale over which 
the condensate wavefunction at the surface 
goes to zero, namely the healing length (\ref{hl}) at the surface. 
To that purpose 
one starts with a spherical trap of radius $R$ such that
$\mu_{TF}=V_{tr}(R)$ (\ref{tfbc}). 
Then by doing a Taylor expansion of $V_{tr}(r)$ about the point $R$ when  
$|R-r| = \Delta \ll R$, one obtains 

\beq
V_{tr}(r) =\mu_{TF} -|\bs{\nabla}V||_{r=R}\Delta + o(2) \label{texpan}\eeq
where the derivative is taken along the direction perpendicular 
to the surface at $r=R$. 
$F$ is the modulus of the attractive external  force 
such that ${\bf F} = - 
\bs{\nabla}V_{tr}$ evaluated at $r=R$. 
Therefore, near the surface the confinement potential takes the form
of a linear ramp potential.
Close to the
boundary,  where $|r-R| \ll R$, the GP equation (\ref{GPEQ}) takes the
form 
\begin{equation}
 - {\hbar^2 \over 2m} {d^2 \over dr^2} \P + (r-R) F
\P +  g  \P^3
= 0\; .
\label{GPB}
\end{equation}
Comparing this equation with the one obtained under  
the usual TF approximation (\ref{chemtf})
we see that the difference comes from 
the inclusion of the term 
due to the  kinetic energy 
(only the most dominant contribution is taken)
as well as the change in the form of the confinement potential.
The surface healing length can now be determined as 
\beq
\delta_s = \left({2 m \over \hbar^2} F \right)^{1/3} \label{b} \eeq
Using this length scale as the unit of the length and scaling the 
wavefunction $\P$ by  
$\delta_{s}(8\pi a)^{1/2}$
the equation (\ref{GPB})  
can be written in a dimensionless form giving 
\beq
\Po^{\prime \prime} - (x + \Po^2) \Po = 0 \; .
\label{rescale}
\eeq
Here $\Po= \delta_s(8 \pi a)^{1/2} \Psi$ and $x=\frac{r-R}{\delta_s}$.
The above equation can be solved to give the surface wavefunction
and the solutions are given in \cite{DLS96}. We mention here the asymptotic
form which gives an exponentially vanishing tail for the condensate 
wavefunction 
\beq
\Po(x \to \infty)  \simeq \frac{A_1}{2x^{1/4}}  \exp
\left( - \frac{2}{3} x^{3/2} \right) \label{plus} \eeq
The Thomas-Fermi solution of $\Psi$ and 
the solution of (\ref{rescale}) determine the behavior of the 
wave function in two distinct regions of space:  the former in 
the interior of the cloud, the latter in the boundary region. 
A full description of the condensate wavefunction is obtained by matching 
these two-type of solutions. The correction  
to the TF approximation is then extended to 
the hydrodynamic theory to obtain the dispersion relation of the surface
modes. This will be discussed now.

\subsection{Surface modes from the hydrodynamic theory}\label{surfhyd}
The harmonic potential can be approximated as a linear ramp potential
in the surface region (\ref{GPB}). Therefore, 
one can reformulate the classical 
hydrodynamic equation for  the collective excitations (\ref{1linhydro})
in the surface of the condensate by replacing the harmonic confinement
with this linear ramp potential \cite{AKPS99} (also
see chapter $7$ of \cite{book}). 
To study the surface modes
one introduces a local two-dimensional 
co-ordinate system in the surface. The $x$ axis of this 
local co-ordinate system is then identified 
with the direction of the gradient of this 
linear ramp potential.
Along the other direction there is no such force and 
hence $\delta \rho$ in this direction  
can be chosen as plane 
wave. We use $\delta_s$ (\ref{b}) as the unit of length
and $\tau=(\frac{2m\hbar}{F^2})^{1/3}$ as the unit of time. Then   
the equilibrium density (\ref{rescale})
has the 
form $\rho =|\Po|^2= -x$ for $x < 0$ and $0$ for $x > 0$
(neglecting the  derivative).  
To obtain the surface dispersion relation one looks for the 
following type of the solutions
\beq \delta \rho = f(qx)e^{qx + iqz} \eeq
The equation  (\ref{1linhydro}) gives  
the dispersion relation  as  
\beq \omega_s^2 = 2q(1+2n), n=0,1, 2, 
\cdots \label{sdisp} \eeq
\beq \delta \rho(x,z,t) = C_{n}L_{n}(-2qx)e^{qx +iqz -i\omega t} \eeq
where $L_{n}(-2qx)$ is the Laguerre polynomial and $C_{n}$ is a constant.
Following \cite{AKPS99}
we shall now state the conditions under which  
this dispersion law for the surface modes
becomes same with the one determined 
(\S\ref{HMTC}) under the global hydrodynamic approximation.
For $l$ much greater than $n$, the dispersion relation (\ref{hydrodisp})
becomes $\omega_{cl}^2=\omega^2 l (1 +2n)$. 
The spherical cloud density has the $l$ dependence of the form 
$P_{l}^{m} (\cos\theta)$ which has the same periodicity of 
$\cos(l\theta)$ (or $\sin(l\theta))$.
The wave number $q$ on the surface of this cloud is therefore given by 
\beq q = \frac{2\pi}{\lambda_q}=\frac{1}{R}\frac{2\pi R}{\lambda_q}
= \frac{l}{R} \label{polarity}\eeq
At the boundary 
the restoring force is 
$F=m \omega^2 R$ (gradient of the harmonic potential)
Eliminating $R$
therefore one gets the agreement  
$\omega_{cl}^2=\omega^2 l(1+2n)=2q(1+2n) = \omega_s^2$ 
(again in the dimensionless unit). 
For $l \gg n$
it is therefore a good approximation to replace 
the parabolic confinement with a linear
ramp potential.  The reason is that the characteristic  penetration depth for 
a mode is of the order 
$\frac{2n+1}{q} = \frac{(2n+1)R}{l}$ from the surface. For $n \ll l$ 
this is much smaller than $R$ and the assumption that these modes 
are confined 
in the surface region is then satisfied.
The agreement is however true for all $l$ when $n=0$ in the 
relation (\ref{hydrodisp}).

In the above derivation only the change in the form of the confinement 
potential from
harmonic to linear ramp is taken into account. 
But the correction due to the 
surface kinetic energy ($\Po^{\prime \prime}$ in Eq.\ref{rescale})
which gives a non
vanishing quantum pressure (\ref{qp}) term in the hydrodynamic
equation is neglected. However 
when the healing length $\delta_s$ (\ref{b})
is not negligible compared to
the wavelength of the surface modes this term is no more negligible
Incorporating the corrections coming from the 
non-vanishing surface kinetic energy through a 
variational approach Khawaja {\it et. al.} \cite{AKPS99} 
obtained the  following dispersion relation  (again in the 
dimensionless unit and for lowest $n$)
\beq \omega_{s}^2 \equiv 2q + 4q^4[-\log 
q  + 0.15] \eeq
This observation agrees with the results by Fetter and Feder \cite{FF98} 
where they have considered the corrections to the Thomas-Fermi description 
of the condensate due to the presence of the boundary layer near the 
condensate surface.

Anglin \cite{JA01} extended this analysis of surface modes 
with small wave-vector $q$ by going beyond Bogoliubov approximation 
through a combination of numerical and analytical techniques. For  
$0 \leq q \leq 2$ the dispersion relation (in  dimensionless form)
turns out to be 
\beq \omega_{s}^2 = 2q + 1.35q^3 + 0.711q^4 \label{anglin}\eeq
Then the ratio  
$\frac{\omega_{s}(q)}{q}$  is minimized to give the 
characteristic rotational frequency of the first vortex
nucleation. 
The critical frequency determines in this way agrees very well 
with the experimentally observed value in MIT\cite{MIT1}
where a strong deformation due to the stirring laser beams can excite 
a surface modes of very large $l$ and the deformed surface region
has a non-negligible width.
For the parameters of MIT experiment using (\ref{polarity}) Anglin
has found  that the mode 
with $l=18$ is unstable against the vortex formation when the 
rotational frequency $\Omega \sim 0.3 \omega_{\perp}$ in an axisymmetric
trap. The agreement is however not so good for the other experiments 
\cite{JILA2, MCWD99, MCBD01, OXFORD}. 
Particularly for the experiment in JILA \cite{JILA2}
where the normal cloud is rotated in a static trap, vortices are nucleated 
before a surface mode is excited. 
For a more detailed discussion on the conditions under which the 
determination of the critical frequency based on Landau criterion 
works well, we refer to \cite{DS00} and \cite{JA01}. In a later chapter  
we shall discuss  a different  theoretical framework to understand 
the vortex nucleation from the surface which is particularly relevant to the 
experiments \cite{MCWD99, MCBD01, OXFORD}. Before that we shall discuss
how the process of vortex nucleation can be studied by applying 
a set of non-local and chiral boundary 
conditions to the linear Schr$\ddot{o}$dinger equation.

\section{Role of boundary conditions in nucleating vortices}
In the earlier sections 
we have discussed the process of vortex nucleation
in the frame-work of repulsively interacting bosons using various 
approximations. In this section we shall provide an alternative 
description of this process where the effect of the interaction can 
be replaced by a set of non-local and chiral boundary conditions
applied to an otherwise non-interacting problem.  
To determine the condensate wavefunction 
$\Psi(\bs{r})$ in a confined geometry like that of 
trapped condensate, one has  
to impose boundary conditions.  
The GP equation (\ref{GPEQ}) is equivalent to a single boson problem in an 
effective one body potential $V_{E}$, given by
\beq V_{E}=V_{tr} + g|\Psi(\bs{r})|^2 \eeq
Because of the presence of the non-linear term $g|\Psi(\bs{r})|^2$, this
equation has to be solved self-consistently with a 
suitable boundary condition \cite{ED95}. The problem becomes 
much simpler if there is a way to replace the non-linear term 
by a suitable choice of the boundary conditions. 
Generally such a
replacement is difficult. This is because  
there is no known mapping available between the 
boundary conditions  and the effective one body potential which 
they aim to replace ( $g|\P(\bs{r})|^2$ in the present problem) 
except for some simple cases.
Therefore the choice of boundary conditions depend on what type of 
effect generated by the effective one body potential the boundary
condition is expected to simulate. 

In the present case we aim to study  
the process of vortex nucleation in a confined 
geometry by solving the linear Schr$\ddot{o}$dinger equation
with a set of such boundary conditions. For simplicity we 
consider a strictly two-dimensional problem.
The proposed boundary conditions are 
motivated by the following consideration. From the previous discussion
we see that the dispersion law for  
surface excitations 
determines the characteristic rotational frequency of vortex 
nucleation. For a two-dimensional problem, equivalent of the 
surface states are the edge states. The proposed boundary condition
therefore should be able to isolate these 
edge states from the bulk states. 
The edge states should have higher angular momentum relative 
to the bulk states. If the proposed
boundary conditions can ensure  this then we
want to see whether with the increasing rotational frequency it is 
energetically favourable for a state with higher 
angular momentum to be transferred from 
the edge to the bulk. This process can then 
be identified with the nucleation of a vortex. 
In the following section 
we shall describe a set of boundary conditions 
which satisfy these conditions. 
Details of this method is given elsewhere \cite{AG03}. 

\subsection{Chiral boundary conditions (CBC)}
We consider the problem of a 
two-dimensional isotropic harmonic oscillator in a co-rotating frame 
rotating  with angular velocity $\Omega$ about the $z$-axis. 
This problem has already 
been discussed in the section \S\ref{2DIO}. There effectively the boundary
conditions are put at infinity. Now 
we seek the solutions of the problem 
in a circular domain of 
radius R. The eigenfunctions and eigenvalues in an infinite domain are already
given in (\ref{2df}) and (\ref{2dre}) of section \S\ref{subsb}.
Using (\ref{hm3}) 
the current density can be derived for a given eigenstate   
\beq  {\bf j} = \frac{\hbar}{2mi}(\Psi_{n,l_z}^{\ast}
\boldsymbol{\nabla} \Psi_{n,l_z} - \Psi_{n,l_z}
\boldsymbol{\nabla} \Psi_{n,l_z}^{\ast} -
2i\frac{m}{\hbar}{\bf A}_{\Omega} |\Psi_{n,l_z}|^2)
\label{current1} \eeq
We have here changed the notation from $\Phi_{\bs{n}}$ to $\Psi_{\bs{n}}$
since they represent the condensate wavefunction and not a single
boson wavefunction. The hamiltonian and the Schr$\ddot{o}$dinger 
equation are however same as those in the single boson problem. 
With the type of eigenfunctions given in (\ref{2df}) it can be checked that
the radial component of such a current density vanishes while its  
azimuthal component is given by   
\beq j_{\theta}=\frac{\hbar}{m}
(\frac{l_z}{r}-\frac{m}{\hbar}\Omega r) |\Psi_{n,l}|^2 \label{current} \eeq
We define $b_{\Omega}=\frac{m\Omega}{h}$. This has the dimension of 
$\frac{1}{L^2}$ and is related to $b_{\omega_\perp}$ through 
$b_{\Omega}=\f b_{\omega_\perp}$ 
 Then with each angular momentum quantum number 
one can associate
a length defined as 
\beq r_{l_z}=\sqrt{\frac{l_z}{b_{\Omega}}} \label{edgerad} \eeq 
For $r < r_{l_z}$, 
$j_{\theta}$ is positive and for $r > r_{l_z}$ it is negative
while it vanishes at $r=r_{l_z}$. 
Using the fact that the hamiltonian (\ref{hm3}) 
is same as that of a charged particle in an effective magnetic field 
$2\Omega \hat{z}$ we call the currents in these two regions  
respectively paramagnetic and diamagnetic. 
For a domain of radius R we also defined quantities
those are similar to the  
magnetic flux in the corresponding Landau problem.
In a dimensionless form 
they are given by  
\beq \fl = b_{\Omega}R^2,~~  \Phi= b_{\omega_\perp}R^2 \label{flux}\eeq

Now let us 
define the bulk and the edge regions 
using this
particular value of $r_{l_z}$ as a reference for a given
angular momentum state such that 
the current associated to that 
particular angular momentum is respectively 
paramagnetic and diamagnetic in  the bulk and the edge. 
Alternatively, one can define the bulk and the edge states for
a disc of size $R$. The bulk states will have angular 
momentum $l_z < b_{\Omega} R^2$ whereas the edge states have
$l_z \ge  b_{\Omega} R^2$. 
We  propose 
a set of non-local and chiral boundary conditions 
for the present problem which   
split the Hilbert space into a direct sum of two orthogonal, 
{\it infinite dimensional} spaces corresponding to bulk
and edge states which respectively have 
positive and negative chirality
on the boundary. The chirality is determined by the direction
of the azimuthal velocity projected on the boundary.
The azimuthal velocity 
 $\frac{j_{\theta}(r)}{|\psi(\vec r)|^2}$ projected on the boundary
of the disc has eigenvalues given by
\beq  \lambda(R)=\frac{1}{R}(l_z
-b_{\Omega}R^2)=\frac{1}{R}(l_z-\Phi_{\Omega}) \eeq
The chiral boundary conditions are  defined in the following way:
\begin{enumerate}
\item For $\lambda \ge 0$,  namely for $0 < \Phi_{\Omega} \le l_z$, 
\beq \partial_r \Psi_{l_z}|_{R}=0 \label{neumann} \eeq
This holds for  any $n$ and henceforth we shall drop the subscript $n$ in 
$\Psi$.

\item For $\lambda < 0$, namely for $l_z < \Phi_{\Omega}$, 
\beq (\frac{\partial}{\partial r}+\frac{i\partial}{r
\partial \theta}+b_{\Omega}r)\Psi_{l_z}|_{r=R}= 0 \label{spectral1} \eeq
\end{enumerate}
For the first set of wavefunctions which accounts for the edge states 
we use Neumann boundary conditions. We could have used as well Dirichlet 
boundary conditions. However unlike Neumann boundary conditions they give an
unphysical discontinuity \cite{Eric1, Narevich}. 
These wavefunctions are more 
and more localized towards the outer 
side of the system with increasing rotational
frequency.
For states with $l_z < \Phi_{\Omega}$, 
whose wavefunctions are localized well inside the disc, we impose
the mixed boundary conditions (\ref{spectral1}). 
These boundary conditions are akin to the boundary conditions
introduced by Atiyah, Patodi and Singer (APS) in their studies 
of Index theorems for Dirac operators with boundaries \cite{APS}.
Similar boundary conditions have also been applied 
to the Landau problem on manifolds with boundaries \cite{Eric1, Narevich}.

\subsection{Spectrum with CBC and the vortex nucleation}
\begin{figure}[h]
\begin{center}
\vspace*{13pt}
\leavevmode \epsfxsize0.75\columnwidth
\epsfbox{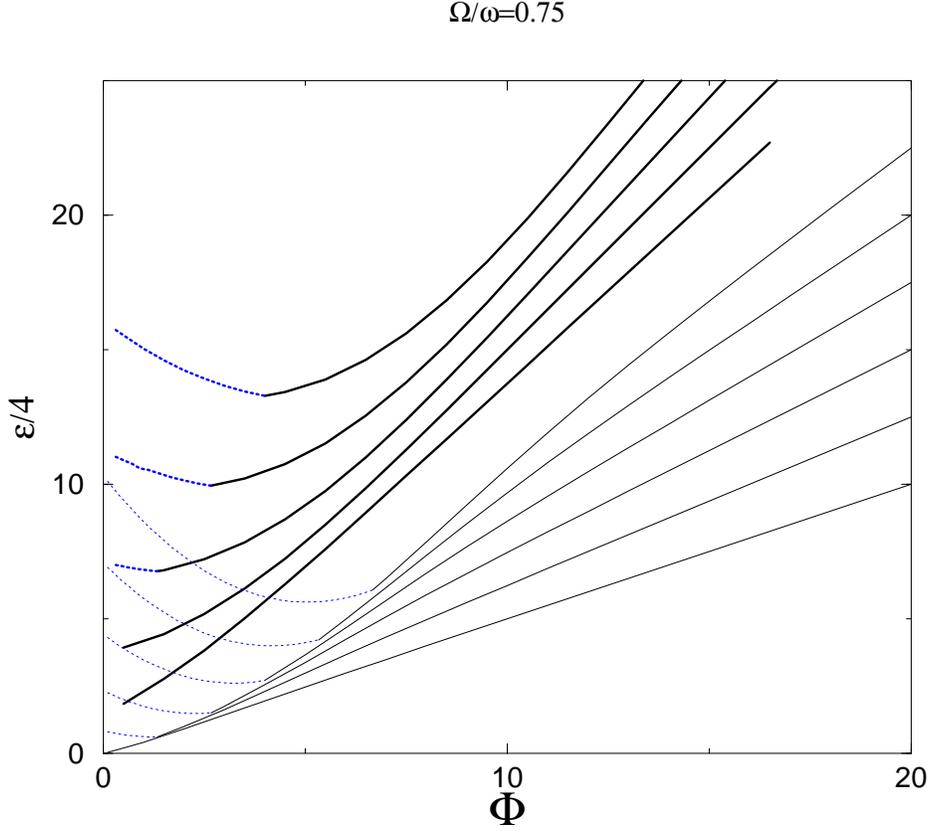}\vskip1.0pc
\caption{\it Energy levels with chiral boundary condition. 
Energy levels for the first few angular momentum
states are shown for $n=0$ and $n=1$. 
Each curve corresponds to a 
given value of the angular momentum.
They correspond to $n=0$ and  
$l_z$ between $0$ and $5$ (thin lines for bulk states and 
thin dotted lines for edge states) as well as $n=1$ 
and $l_z$ between $-1$ and $3$ with $l_z=-1$
corresponds to the lowest curve (thick lines for bulk states
and thick dotted lines for the edge states). 
For $l_z > 0$ 
the spectrum has
a kink at the point $l_z=\Phi_{\Omega} =\f \Phi$.}
\label{fig2}
\end{center}
\end{figure}

To describe the spectrum let us introduce the dimensionless form
of the energy, namely 
\beq \varepsilon=\frac{2mE_{n,l_z}R^2}{\hbar^2} = 
(\frac{2 E_{n,l_z}}{\hbar \omega_{\perp}})\Phi \label{dimen} \eeq
where $\Phi$ is defined in (\ref{flux}).
According to the chiral boundary conditions when $\Phi$ is increased
at a fixed $\f$, the sign of the eigenvalues 
$\lambda(R)$ changes from  positive to 
negative. Correspondingly the energy $\varepsilon$ (\ref{dimen}) of 
a state with a given $n$ and $l_z$ changes. This change in energy 
describes the corresponding transfer of a state   
from the edge to the bulk Hilbert spaces at the point 
$\Phi_{\Omega}=l_z$ (\ref{neumann}-\ref{spectral1}).
as shown in Fig.\ref{fig2}.   
For large $\Phi$, the infinite plane solutions (\ref{2de}) are reached
asymptotically.

\subsection{Spectrum with CBC}

For an infinite system we have $\frac{\varepsilon}{4\Phi} = 
n+\frac{1}{2}(1+(1-\f)l_z)$. 
Therefore $\frac{\varepsilon}{4\Phi}$, for a given $\f$, 
is a linear function of $l$ with slope $(1-\f)$.
When chiral boundary conditions are applied, 
this behaviour is approximately obeyed for
the bulk states. But for the edge states the energy increases
non-linearly with increasing angular momentum.

\begin{figure}[h]
\begin{center}
\vspace*{13pt}
\leavevmode \epsfxsize0.75\columnwidth
\epsfbox{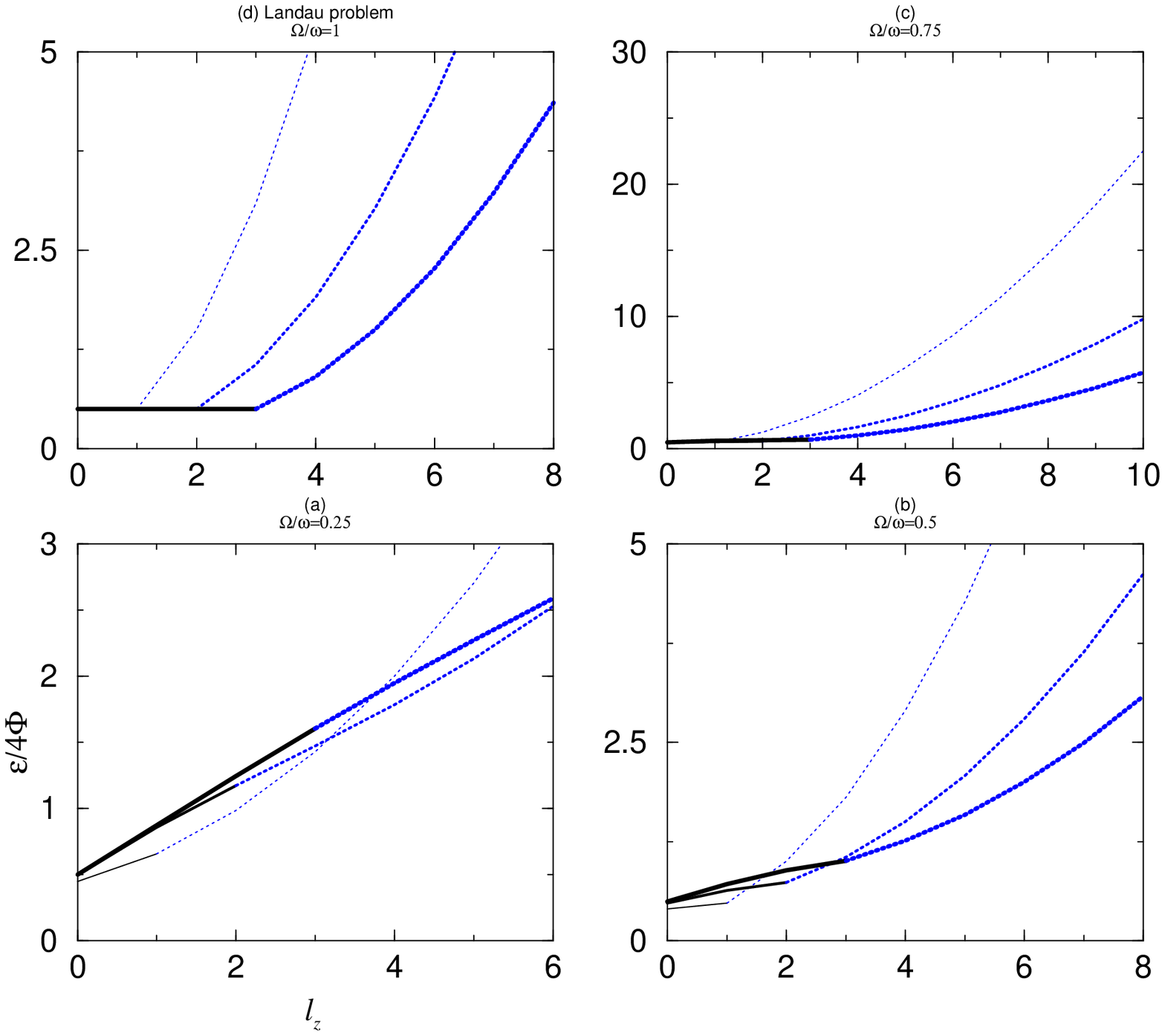}\vskip1.0pc
\caption{\it Effect of a faster rotation. 
In these figures we have plotted $\frac{\varepsilon}{4\Phi}$ 
as function of $l_z$ for a set of $\f$ values (given above each figure). 
The three set 
of plots in each figure correspond to $\Phi_{\Omega}=1,2,3$
(the thinnest one for $\fl=1$ and the thickest one for $\fl=3$). 
The dotted part corresponds to edge states while the continuous part
corresponds to bulk states. For $\f=1$ bulk states for all three 
values of $\fl$ fall on the same line.}
\label{fig3}
\end{center}
\end{figure}

The Fig.\ref{fig3} shows the effect
of an increase of the  rotational frequency 
on the spectrum under the choice of CBC.  
We have plotted 
the energies of the bulk and the edge states 
for four different values
of $\f$. For each value of 
$\f$,  the quantity $\Phi_{\Omega}$ 
is increased by unit steps from
$1$ to $3$ and the corresponding bulk and edge energies are shown.
Under these conditions, 
the slope of the bulk energy levels increases while the slope
of the edge energy levels goes down.
The opposite behaviour is observed when, for a fixed $\fl$, the ratio $\f$
increases. 
There are therefore certain values of the ratio $\f (< 1)$, 
at which the edge states for
a given $\fl=l_z$ just intersect the bulk state corresponding to
$\fl=l_z+1$. The first vortex is nucleated for $l_z=1$. 
The characteristic rotational frequency is denoted by $\Omega_{c1}$.
The rotational frequency at which the edge states for 
$\fl=1$ intersects the bulk states 
of the $i$-th vortex with $l_z=1$ is denoted by $\Omega_{ci}$.

A finite Bose-Einstein condensate without any vortex 
is defined when the 
domain of radius $R$ (bulk region)
contains only the state $l_z=0$
such that $R = r_{1}$ (\ref{edgerad}). 
The boundary is given by $\fl=1$
and the corresponding bulk and edge energy levels are shown as a 
function of the ratio $\f$ in the Fig.\ref{fig3}. 
When $\fl=2$ instead for
$\fl=1$, the state
with $l_z=1$ is transferred 
from the edge to the bulk Hilbert space.
For $\Omega \ge \Omega_{c1}$ 
the energy 
$\frac{\varepsilon}{4\Phi}(l_z)$ (\ref{dimen}) of any state with $l_z \ge 2 $
is less if  $\fl=2$ 
instead of $\fl=1$. 
At $\Omega=\Omega_{c1}$ the boundary is given by 
$\fl=2$. The state with $l_z=1$ 
is now transferred to the bulk Hilbert space at this 
characteristic frequency. 
The bulk region has now one unit($\frac{h}{m}$) of extra  
rotational flux. 
Thus for  $\Omega=\Omega_{c1}$
a vortex with $\kappa=1$ is  nucleated in the bulk.
At this characteristic value for $\Omega$ 
the region with $R=r_{2}$ (\ref{edgerad})
is defined as a condensate with a single vortex of $l=1$. 

From these boundary conditions we can thus 	
determine the characteristic 
frequency of nucleation of the first
vortex in terms of the trap frequency. 
It lies in between  $\Omega=0.35\omega_\perp$ and $\Omega=
0.36\omega_\perp$. We note
that this  characteristic frequency 
of nucleation is close to the value
$\Omega=
0.29 \omega_\perp$ that has been observed  
in one of the experiments
\cite{MIT1} where $\omega_\perp$ is the trap frequency in the transverse direction. 

\subsection{Vortex nucleation at $\Omega > \Omega_{c1}$}
We have so far  
discussed only the the nucleation of the first vortex with $\kappa=1$ and
we have determined the characteristic rotational frequency. With a further
increase of the  rotational frequency 
the boundary is successively given by $\fl = 3, 4, 5, 
\cdots $. Correspondingly more than one higher angular momentum states are
transferred from the edge to the bulk Hilbert space at a time. 
The characteristic  rotational 
frequencies are respectively given by  
$0.42 <\frac{\Omega_{c2}}{\omega_\perp} <0.43$, 
$0.46 < \frac{\Omega_{c3}}{\omega_\perp}  < 0.47$, $0.48< 
\frac{\Omega_{c4}}{\omega_\perp}
<0.49$. For example, 
if the rotational  frequency of the trap is ramped up
to $\Omega_{c2}$ 
and the system is allowed to come to equilibrium in the co-rotating frame, 
the boundary is given by $\fl=3$ instead
of $\fl=1$. Since this is accompanied by the transfer of two units of the 
rotational flux quanta to the bulk, either two
vortices each with $\kappa=1$ are nucleated or a single vortex 
with $\kappa=2$ is nucleated. 
The first of these two alternative situations  implies that at   
$\Omega=\Omega_{ci}$, $i$ number of  
vortices with $\kappa=1$ enters. 
Alternatively the rotational frequency may be increased adiabatically
from $0$ to a higher value. 
The first vortex will enter at $\Omega_{c1}$.
The second vortex with $\kappa=1$ enters when the bulk states
for $\fl=3$ just intersects 
edge states for $\fl=2$. The corresponding rotational frequency  
is relatively higher. 

Some features of the vortex nucleation can thus be explained by 
replacing the non-linear interaction with 
the chiral boundary conditions. Quantitative comparison of 
the theory with experimental predictions is difficult
since the experimental geometry is generally more complicated
than a pure two-dimensional disk. 
Particularly we have not considered the deformation of the  
the trap from the initial circular shape in the plane of the rotation
\cite{MCWD99, MCBD01, MIT1, OXFORD}. 
This deformation effects the process of the vortex nucleation in 
these rotating traps. In the following
section we shall briefly describe how this deformation is incorporated 
in the theoretical description of the problem of vortex nucleation.

\section{Vortex nucleation in a deformed trap}\label{detrap}
It is mentioned in section \S\ref{rotdeform} that  
to rotate the trap an  anisotropy (\ref{anisotropy}) is introduced 
in the confinement potential by a laser stirrer. This stirrer is then 
rotated to 
create the rotating harmonic trap. Subsequently, beyond a characteristic
rotational frequency 
the first vortex is nucleated. In this section we discuss 
the role of this anisotropy in the vortex nucleation. 
To that purpose, following \cite{RZS01} 
we start by considering the steady state solutions 
of the hydrodynamic equations (\ref{rothydr1}-\ref{rothydr2}) in 
such rotating traps. 
\subsection{Stationary states in the rotating frame} 
It is already known from classical hydrodynamics that for
a uniform fluid in a rotating elliptical cylinder, the 
instantaneous induced velocity field in the lab frame is 
given by \cite{fsr01, Lamb45}
\beq  \bs{v}_{cl}=\Omega \frac{A^2 - B^2}{A^2 + B^2}\bs{\nabla}(xy) \eeq
where $A$ and $B$ are the major and the minor axis of the ellipse.
In order to  solve the equations (\ref{rothydr1}-\ref{rothydr2}
one may therefore propose an  ansatz 
velocity field of the following form  
\beq \bs{v}_s \propto \alpha \bs{\nabla}(xy) \eeq  
Here $x$ and $y$ are now the co-ordinates in the co-rotating frame
(identical as  $X$ and $Y$ in \S\ref{rotdeform}).
The corresponding steady-state 
ellipsoidal density can be written in the usual TF 
form (\ref{TFDP})
\beq \rho(\bs{r})=\frac{1}{g}
\left[\mu_{TF} - \frac{m}{2}(\tilde{\omega}_x^2
x^2 + \tilde{\omega}_y^2 y^2 + \omega_z^2 z^2 \right]\eeq
with 
\bea \tilde{\omega}_x^2 &=& \omega_x^2 + \alpha^2 - 2\alpha \Omega \nonumber \\
 \tilde{\omega}_y^2 &=& \omega_y^2 + \alpha^2 + 2\alpha \Omega \eea
To see why the above form of the steady state density 
corresponds to a quadrupole
mode let us rewrite it as 
\beq \rho(\bs{r})=\frac{1}{g}\left[\mu_{TF} - \frac{m}{2}\left(
(\omega_{\perp}^2 + \alpha^2)r^2 + \omega_z^2 z^2 \right)\right]  
-\frac{1}{g}[\frac{m}{2}(\epsilon - 2 \alpha \Omega)(x^2 - y^2)] 
\label{quaddens}\eeq
In terms of the co-ordinates in the co-rotating frame, the steady 
state density is therefore a sum of the usual TF density in a static trap
(after replacing $\omega_\perp^2$ with $\omega_\perp^2 + \alpha^2$)
and a quadrupolar density. 
The deformation of the  of the trap is given by 
a quantity $\delta$ such that 
\beq \delta = \frac{\tilde{\omega}_x^2 - \tilde{\omega}_y^2}
{\tilde{\omega}_x^2 - \tilde{\omega}_y^2} = 
\frac{\epsilon - 2\alpha \Omega}{1 + \alpha^2} \label{defm}\eeq
The continuity equation (\ref{rothydr1}) requires that 
\beq \alpha = -\delta \Omega \eeq

These equations are satisfied simultaneously provided 
\beq \alpha^2 + \alpha (\omega_{\perp}^2 - 2 \Omega^2) + \Omega \epsilon
\omega_{\perp}^2 =0 \eeq

Solution of the above
equation gives the stationary solutions in the  
$\Omega$, $\alpha $ plane and has been analyzed in detail in 
\cite{RZS01}. We mention some special features. For a
symmetric trap potential ($\epsilon=0$) if  
$\Omega < \frac{\omega_\perp}{\sqrt{2}}$, $\alpha =0$. Substituting this
in (\ref{quaddens}) gives the usual TF density of the cloud in an  
axisymmetric trap (\ref{TFDP}). However for 
$\Omega < \frac{\omega_\perp}{\sqrt{2}}$, the equation gets a new class
of solutions which correspond to 
$\alpha = \pm \sqrt{2\Omega^2 - \omega_\perp^2}$. This 
non-zero value of $\alpha$ for $\epsilon =0$ corresponds to 
a spontaneous deformation of the cloud to a quadrupolar shape. 
The physical
origin of this spontaneous deformation of the trap is as follows. 
With the increase in the rotational frequency the energy of the quadrupolar
state with $l_z = +2$
monotonically decreases to $0$ till $\Omega =\omega_{\perp}$. Beyond this
value of $\Omega$ the energy of this mode becomes negative. 
This leads to a thermodynamic instability
and the system is spontaneously deformed at this $\Omega$. 
For non-zero $\epsilon$, the phase diagram corresponds to 
two branches of solutions known respectively as, the normal branch and the 
over-critical branch. We do not detail  these results any further and now
go to the issue of vortex  nucleation in such a deformed trap.

\subsection{Adiabatic switch on of the rotation}
The above mentioned solutions are  stationary solutions of
the hydrodynamic equations in the rotating frame. So, to produce them
experimentally the rotation has to be ramped up adiabatically for 
a fixed anisotropy $\epsilon$.
In the corresponding experiments \cite{CMBD01, OXFORD} it is found
that at higher rotational frequency which is generally
around $\Omega = \frac{\omega_{\perp}}{\sqrt{2}}$, these stationary 
solutions show dynamic instability leading to the vortex nucleation.
By performing a linear stability 
analysis of the equations of rotational hydrodynamics 
Sinha and Castin \cite{CS01} showed that this is 
indeed the case. 
They found that if $\epsilon$ is very close to $0$, 
at $\Omega=\frac{\omega_{\perp}}{\sqrt{2}}$, a large quadrupole oscillation
makes the stationary solution dynamically unstable. For larger $\epsilon$
the dynamic instability occurs
over a larger range of $\Omega$ around $\Omega=
\frac{\omega_{\perp}}{\sqrt{2}}$. 
Solving the time dependent GP equation (see also \cite{CBB99}) they also
verified that this dynamic instability leads to the vortex nucleation.
Their findings are particularly in agreement with the experimental results in 
\cite{CMBD01}. 

\subsection{ Sudden switch on of the rotation}
The early ENS experiments \cite{MCWD99, CMD00} and the MIT experiments 
\cite{MIT1, MIT2}
come under this class. Here the stirring potential is rotated at a 
fixed frequency $\Omega$ and the deformation (ellipticity) due to 
the stirrer is 
switched on suddenly to a given value of $\epsilon$. In the MIT experiment
\cite{MIT1}
the deformation is large  and consequently it excites high polarity 
surface modes. We have already discussed in (\S\ref{surfhyd})
how the characteristic 
nucleation frequency can be obtained in this case \cite{JA01}
by using the Landau criterion. However in the 
other experiments the deformation is relatively less and the 
vortex nucleation can be 
understood by studying the quadrupole modes. 
Here we discuss the vortex nucleation mechanism following 
Kraemer {\it et. al.}\cite{KPSZ01}. Their study involves
the evaluation of the energy of 
a displaced vortex \cite{fsr01, SF00L, GG01}
in the presence of 
a surface quadrupolar deformation under the 
sudden switch on of the trap deformation. 

We start by evaluating the extra energy and the angular momentum
associated with 
the presence of a 
straight vortex line with $\kappa=1$
in a perfectly cylindrical
axisymmetric trap (axis of symmetry is
again the $z$-axis) displaced from the center of the cylinder by 
a distance $d$. 
The corresponding expressions in TF approximation are respectively

\beq
L_z(d/R_{\perp})=
N_0\hbar\left[1-\left(\frac{d}{R_{\perp}}\right)^2\right]^{5/2}\,,
\label{amd}
\eeq
and 
\beq
E_v(d/R_{\perp},\mu_{TF})=
E_v(d=0,\mu_{TF})\left[1-\left(\frac{d}{R_{\perp}}\right)^2\right]^{3/2}\,,
\label{epv}
\eeq
where $E_v(d/R_{\perp}, \mu_{TF})$ is given in (\ref{lundven}).
According to (\ref{amd}) and (\ref{epv}) if the vortex is exactly located 
at the boundary it neither carries extra angular momentum nor 
energy relative to a vortex free condensate. 
When the trap is rotated with angular velocity $\Omega$ about the $z$-axis 
the energy of the system in the co-rotating frame 
is  given by 

\beq
E_v(d/R_{\perp},\Omega,\mu_{TF}) = 
E_v(d=0,\mu_{TF})\left[1-\left(\frac{d}{R_{\perp}}\right)^2\right]^{3/2}-
\Omega N_0\hbar\left[1-\left(\frac{d}{R_{\perp}}\right)^2\right]^{5/2}\,.
\label{Evdomega}
\eeq   
From this expression one finds that 
a vortex at $d=0$ is energetically favorable
if the rotational frequency satisfies
$\Omega \ge \Omega_{c1}(\mu_{TF}) = E_v(d=0,\mu_{TF})/N_0\hbar$
which is the thermodynamic criterion of vortex stability (\S\ref{gpcritf}). 
However, even for  $\Omega\ge\Omega_v(\mu)$ and hence
if $E_v(d/R_{\perp},\Omega,\mu)$ is negative at $d=0$, the energy
exhibits a maximum at intermediate values of $d$
 that lies between $0$ and $R_{\perp}$. Therefore if the rotation is suddenly 
switched on at a value greater than $\Omega_{v}$, because of this 
energy barrier a vortex cannot be nucleated into the center of the trap
though the vortex 
state is a global minima of the energy.
To summarize the situation, 
the energy landscape will look like a valley representing this 
global minima surrounded by 
this energy barrier.
As suggested by the experiments in ENS, MIT and Oxford, 
the disappearance of the barrier at sufficiently high angular 
velocity is preceded by a quadrupolar deformation of the atomic cloud.
Therefore this energy landscape changes in the presence of a quadrupolar
mode. In an elliptically deformed trap
the energy of a quadrupole deformation in the rotating frame is given by
\cite{RZS01}  
\beq 
E_Q(\delta,\f,\epsilon,\mu_{TF}) =N_0\mu_{TF}
\left[(\frac{2}{7})\frac{1-\epsilon\delta-
(\f)^2\delta^2}{\sqrt{1-\epsilon^2}\sqrt{1-\delta^2}}+
\frac{3}{7}\right]\,.
\label{EQ}
\eeq
In \cite{KPSZ01}
using this expressions the total energy of a quadrupole 
deformed condensate in presence of a 
displaced vortex can be evaluated. The energy
landscapes are then redrawn at various values of the rotational frequency 
to understand the role of quadrupole deformation on the previously
described energy barrier under sudden switch on of the deformation
of the trap. At the time of the sudden switch-on of the deformation,
$\delta=0$. This corresponds to $\epsilon =0$. Then the system will
evolve through a set of intermediate configurations before 
attaining a stable value of the deformation 
parameter $\delta$. By noting the corresponding changes
in the energy barrier it is found \cite{KPSZ01} that at a given 
value for the trap deformation $\delta$
the  a saddle is created in the energy barrier 
such that the vortex energy along
that path continually decreases from the surface to the center.
At this point the vortex nucleates. This explains the vortex nucleation
mechanism under the sudden switch on of the trap deformation.

\section{Conclusion and Outlook}
In this article we have discussed the problem of vortex nucleation in a 
finite quantum system, namely the trapped 
atomic condensate. We have described how the 
thermodynamic description of the vortex nucleation needs to be 
modified in a finite trap geometry and the explanation of vortex 
nucleation requires 
a detailed analysis of the condensate at its surface 
or edge (for a two-dimensional geometry). Though the local theory 
takes different form to analyze the experiments based on different
mechanism to rotate the trap, the unifying aspect of these approaches 
is the role played by surface Bogoliubov excitations. 
There are 
many issues related to the nucleation of vortices in BEC that have not 
been discussed in this article.  
They include for 
instance the problem of vortex
nucleation in attractive condensate, the dynamics and evolution 
of the vortex lattice state and the nature of the many-vortex state
in a rapidly rotating BEC. All these problems constitute an
active domain of research both 
on experimental and theoretical fronts. 
Also for the topic discussed,  
the choice of the references which we
have selected for our discussion is not all-inclusive due to the limitation of
space. We take this opportunity to apologize for what is left out and
to mention that this by no means diminishes their importance.
For a more complete account we again 
refer to the review by Fetter \cite{fsr01} and the references
therein.

\vskip 20pt
\noindent{\large\bf Acknowledgements}\\\\
This work primarily grows out of a collaboration with Prof. Eric Akkermans
to whom I owe much of my understanding about vortices in BEC. 
I am very grateful to him for a very careful reading of the manuscript. 
I am also very thankful
to Prof. Assa Auerbach for many discussions on vortex physics and particularly
on Bogoliubov excitations. I would also take this opportunity to thank Dr. 
Subhasis Sinha for many general discussions on rotating condensates.
This work is supported by the Israel 
Council for Higher Education, the Technion, the  Israel Academy of Sciences 
and the fund for promotion of Research
at the Technion.

\end{document}